\documentclass[prd,nofootinbib,12pt]{revtex4}

\usepackage{lineno,hyperref}
\usepackage{amsfonts,amssymb,amsmath}
\usepackage{epsfig}
\usepackage{mathrsfs}
\usepackage{amssymb}
\usepackage{graphicx}
\usepackage{amsmath}
\usepackage{graphicx}
\usepackage{amsmath}
\usepackage{xcolor}
\usepackage{color}
\usepackage{appendix}

\begin{document}

\title{A quantum information method for early universe with non-trivial sound speed}

\author{Shi-Cheng Liu$^{1}$}
\author{Lei-Hua Liu$^{1}$}
\email{liuleihua8899@hotmail.com}
\author{Bichu Li$^{1}$}
\email{libichu@mail.ustc.edu.cn}
\affiliation{$^1$Department of Physics, College of Physics, Mechanical and Electrical Engineering, Jishou University, Jishou 416000, China}
\author{Hai-Qing Zhang$^{2}$}\email{hqzhang@buaa.edu.cn}
\affiliation{Center for Gravitational Physics, Department of Space Science, Beihang University, Beijing
	100191, China}
\affiliation{Peng Huanwu Collaborative Center for Research and Education, Beihang University, Beijing 100191, China}
\author{Peng-Zhang He$^{3}$}\email{hepzh@buaa.edu.cn}
\affiliation{School of Physics and Astronomy, China West Normal Uniersity, Nanchong 637002, Sichuan, China}

\begin{abstract}

Many quantum gravitational frameworks, such as DBI inflation, k-essence, and effective field theories obtained by integrating out heavy modes, can lead to a non-trivial sound speed. Meanwhile, our universe can be described as an open system. Under the non-trivial sound speed, we employ the method of open quantum systems combined with Arnoldi iterations to study the Krylov complexity throughout the early universe, including the inflationary, radiation-dominated, and matter-dominated epochs. A key ingredient in our analysis is the open two-mode squeezed state formalism and the generalized Lanczos algorithm. To numerically compute the Krylov complexity, we are the first time to derive the evolution equations for the parameters $r_k$ and $\phi_k$ within an open two-mode squeezed state. Our results indicate that the Krylov complexity exhibits a similar trend in both the standard case and the case with non-trivial sound speed. To distinguish between these two scenarios, we also investigate the Krylov entropy for completeness. The evolution of the Krylov entropy shows a clear difference between the standard case and the non-trivial sound speed case. Furthermore, based on the behavior of the Lanczos coefficients, we find that the case of non-trivial sound speed behaves as a maximally chaotic system. However, our numerical results suggest that the Krylov complexity does not saturate to a constant value due to the huge expansion of spacetime background. This study offers a new perspective for exploring the early universe through the quantum information.

\end{abstract}

\maketitle


\section{introduction}
\label{introduction}

The curvature perturbation serves as the seed for large-scale structure formation, yet its origin remains enigmatic. Although the Bunch-Davies (BD) vacuum is widely regarded as a suitable state for explaining observations of the cosmic microwave background (CMB) \cite{Agullo:2022ttg}, it is generally impossible to define a privileged quantum vacuum state in curved spacetime \cite{Unruh:1976db}. In this context, the BD vacuum can be viewed as an approximation to the Minkowski vacuum at short distances. This suggests that the origin of the curvature perturbation’s vacuum may lie at a deeper level. As pointed out in \cite{Grishchuk:1990bj}, the amplification of quantum fluctuations to classical scales is naturally a squeezing process. Consequently, the two-mode squeezed state has become a vital concept in inflationary cosmology.

The growing application of two-mode squeezed states in cosmology underscores the increasing relevance of quantum information concepts across various domains. Among these, Krylov complexity, a measure of operator growth in quantum dynamical systems based on the Lanczos algorithm, has garnered significant attention \cite{Parker:2018yvk}. It has been widely applied in areas such as the SYK model in high-energy physics \cite{Rabinovici:2020ryf,Jian:2020qpp,He:2022ryk}, the Ising and Heisenberg models \cite{Cao:2020zls,Trigueros:2021rwj,Heveling:2022hth}, conformal field theory \cite{Dymarsky:2021bjq,Caputa:2021ori,Kundu:2023hbk}, quantum field theory \cite{He:2024hkw,He:2024xjp,He:2025guu}, and topological phases \cite{Caputa:2022eye}. In generic settings, Krylov complexity exhibits exponential growth, even in integrable systems, as verified in \cite{Camargo:2022rnt,He:2024xjp}. Moreover, Krylov complexity has been extended to open quantum systems \cite{Bhattacharyya:2025lsc,Zhai:2024tkz,Carolan:2024wov,Bhattacharya:2023zqt,Liu:2022god,Bhattacharjee:2022lzy}. It also plays a role in the holographic dictionary \cite{Li:2025fqz,Aguilar-Gutierrez:2025kmw,Heller:2024ldz,Das:2024tnw,Rabinovici:2023yex}, with notable connections to JT gravity \cite{Bhattacharyya:2025gvd,Ambrosini:2024sre} and dilaton physics \cite{Chattopadhyay:2023fob}. For the most recent developments in this rapidly evolving topic, we refer readers to Refs. \cite{Rabinovici:2025otw,Nandy:2024evd}. Essentially speaking, the study of complexity in the early universe is fundamentally an investigation into quantum information within a non-inertial framework. Because the expansion of spacetime renders observers non-inertial relative to distant cosmological scales, it is crucial to consider the perspective of quantum information in non-inertial systems. This approach aligns with recent developments in relativistic quantum information, such as those discussed in \cite{Wu:2022xwy,Wu:2023spa,Wu:2023sye,Li:2025bzd,Li:2025jlu,Liu:2024wpa,Liu:2025hcx}.

In this work, we will employ the framework of Krylov complexity to study the early universe, encompassing inflation, the radiation-dominated (RD), and matter-dominated (MD) eras, within models featuring a non-trivial sound speed. While several previous studies have explored computational complexity during inflation \cite{Choudhury:2020hil,Bhargava:2020fhl,Lehners:2020pem,Bhattacharyya:2020rpy,Adhikari:2021ked}. The work of \cite{Adhikari:2022oxr} represents the first attempt to apply Krylov complexity to inflation, though it relies on a closed-system approach. In reality, however, the early universe constitutes an open quantum system. Recently, \cite{Li:2024kfm} adopted an open-system approach to construct the wave function of the inflationary universe, revealing its strongly dissipative nature. Building on this perspective, the present study takes the wave function in open systems and the generalized Lanczos algorithm as its foundational starting point.

In our previous studies, we examined the Krylov complexity of the thermal state across inflation, radiation-dominated (RD), and matter-dominated (MD) eras using the simplest inflationary potential, the quadratic potential \cite{Li:2024ljz,Li:2024iji}. Meanwhile, Ref. \cite{Zhai:2024odw} has demonstrated that different inflationary potentials yield nearly identical evolution for both Krylov complexity and Krylov entropy when preheating dynamics are taken into account. This implies that, within the single-field inflation framework, various potentials cannot be distinguished based on these measures. Since a non-trivial sound speed can also give rise to diverse inflationary scenarios, this work aims to investigate the Krylov complexity in such settings, in order to differentiate them from standard single-field inflation. 

This paper is organized as follows. In Sec. \ref{section some basics of early universe}, we provide foundational background on the early universe and the concept of non-trivial sound speed. Section \ref{the hamiltonian} briefly presents our target Hamiltonian. In Sec. \ref{lanczos algorithm}, we introduce the generalized Lanczos algorithm based on the Arnoldi iteration and discuss the statistical properties related to the Lanczos coefficient. Section \ref{krylov complexity and krylov entropy} examines the Krylov complexity and Krylov entropy using numerical results for $r_k$ and $\phi_k$. Finally, Sec. \ref{summary and discussion} offers conclusions and further discussion.


\section{Some setup of early universe}
\label{section some basics of early universe}

Building upon the framework established in \cite{Zhai:2024odw}, we model the scale factor evolution across key cosmological epochs:
\begin{itemize}
	\item \textbf{Inflation}: Exponential expansion driven by scalar field dynamics
	\item \textbf{Radiation domination (RD)}: $a(t) \propto t^{1/2}$ ($t$ is the physical time)
	\item \textbf{Matter domination (MD)}: $a(t) \propto t^{2/3}$
\end{itemize}

While slow-roll conditions typically suppress the influence of inflationary potentials inflation, \textit{violations of slow-roll} necessitate explicit inclusion of potential contributions. These violations occur during the preheating phase, where potential energy converts to particle production via the process:
\begin{equation}
V(\phi) \rightarrow \text{kinetic energy} \rightarrow \text{particle creation}
\end{equation}

This particle production mechanism provides the initial conditions for analyzing Krylov complexity ($\mathcal{C}_K$) in subsequent RD and MD eras. Crucially, \cite{Zhai:2024odw} demonstrated that for $\mathcal{C}_K$ and associated entropy measures during RD/MD:
\begin{equation}
\Delta\mathcal{C}_K^{V_1} \approx \Delta\mathcal{C}_K^{V_2} \approx \Delta\mathcal{C}_K^{V_3}
\end{equation}
for diverse potentials $V_i(\phi)$. This \textit{universality} justifies our focus on the minimal quadratic potential:
\begin{equation}
V(\phi) = \frac{1}{2}m^2\phi^2
\end{equation}
as a representative case capturing essential features while simplifying calculations.

\subsection{Inflation, RD and MD}
Following Ref. \cite{Baumann:2009ds}, we first introduce the so-called Friedman-Lemaitre-Robertson-Walker (FLRW) metric as follows, 
\begin{equation}
ds^2=a^2(\eta)(-d\eta^2+d\vec{x}^2),
\label{bacground metric}
\end{equation}
where $a(\eta)$ is the scale factor. For simplicity, we only consider one single component universe, indicating the universe only experiences inflation, RD and MD. We consider a universe dominated by a single component at each epoch, transitioning sequentially through inflation, RD, and MD. The equation of state parameter characterizing each epoch is defined as
\begin{equation}
w_I = \frac{P_I}{\rho_I},
\label{eq:eos}
\end{equation}
where $P_I$ and $\rho_I$ denote the pressure and energy density in epoch $I$ (inflation, RD, or MD).

To establish the relation between conformal time $\eta$ and scale factor $a$, we consider the comoving particle horizon:
\begin{equation}
\chi_{\mathrm{ph}}(\eta) = \int_{t_i}^{t} \frac{\mathrm{d}t'}{a(t')} = \int_{\ln a_i}^{\ln a} (aH)^{-1} \mathrm{d}\ln a',
\label{eq:comoving_distance}
\end{equation}
where $\chi_{\mathrm{ph}}(\eta)$ represents the comoving distance to the particle horizon. The Hubble radius $(aH)^{-1}$ can be expressed in terms of the equation of state parameter as
\begin{equation}
(aH)^{-1} = H_{0}^{-1} a^{\frac{1}{2}(1 + 3w)},
\label{eq:hubble_radius}
\end{equation}
where we simplify notation by using $w$ for $w_I$. Substituting \eqref{eq:hubble_radius} into \eqref{eq:comoving_distance} and integrating yields the conformal time-scale factor relation:
\begin{equation}
\eta = \frac{2H_{0}^{-1}}{1 + 3w} a^{\frac{1}{2}(1 + 3w)} + C,
\label{eq:eta_a_general}
\end{equation}
where $C$ is an integration constant determined by boundary conditions. For specific cosmological epochs with characteristic $w$ values, we obtain:
\begin{equation}
\eta=\begin{cases}
& -(aH_{0})^{-1},  \   \ (\omega=-1), \ \  \rm inflation \\
& aH_{0}^{-1}, \  \ (\omega=\frac{1}{3}), \ \ \rm RD \\
& 2\sqrt{a}H_{0}^{-1}, \  \ (\omega=0), \ \ \rm MD
\end{cases}
\label{formula of eta}
\end{equation}
These solutions demonstrate how distinct values of $w$ govern the $\eta$-$a$ relationship in different cosmological eras.

\subsection{Non-trivial sound speed during different epochs}
\label{nontrivial sound speed in different epochs}

The non-trivial sound speed will lead to the non-trivial kinetic term, represented by $c_S$. When $c_S^2=1$, the kinetic term will become the standard canonical kinetic term. The generic non-trivial sound speed can be defined in k-inflation in terms of Mukhannov-Sasaki variable \cite{Armendariz-Picon:1999hyi,Garriga:1999vw}. And the non-trivial sound speed can also be embedded into the string framework \cite{Alishahiha:2004eh,Silverstein:2003hf}. Meanwhile, the non-trivial sound speed can also be realized in DBI inflation \cite{Chen:2020uhe}. The effective field theory can also obtain the non-trivial sound speed after integrating the heavy modes in the IR region \cite{Achucarro:2012sm}. Thus, we could see that many quantum gravitational frameworks could lead to the so-called non-trivial sound speed. In this work, we will adopt the so-called sound speed resonance \cite{Cai:2018tuh} (SSR) as follows,
\begin{equation}
\begin{split}
c_S^2=1-2\xi[1-\cos{k\eta}]
\end{split}
\label{ssr}
\end{equation}
where $\xi$ is the amplitude of oscillation and $k$ is the momentum. This SSR can be realized in DBI and effective field theory \cite{Chen:2020uhe,Chen:2019zza}, $\it etc. $. Thus, one could see that the SSR formula \eqref{ssr} is of significant meaning in the quantum gravitational framework. However, we will relax the constrained condition for \eqref{ssr}, allowing the SSR to be applied to inflation, RD, and MD.

\begin{figure}
	\centering
	\includegraphics[width=0.45\linewidth]{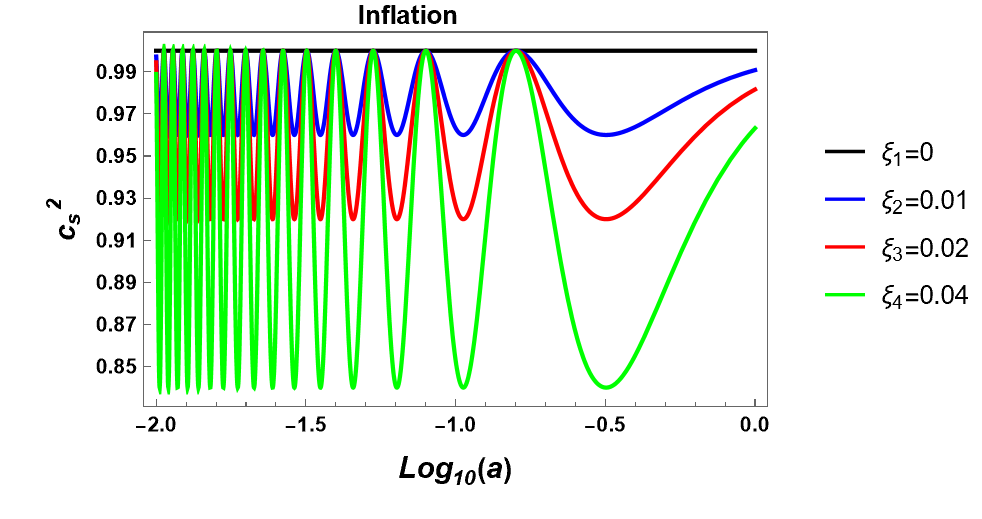}
	\qquad
	\includegraphics[width=0.45\linewidth]{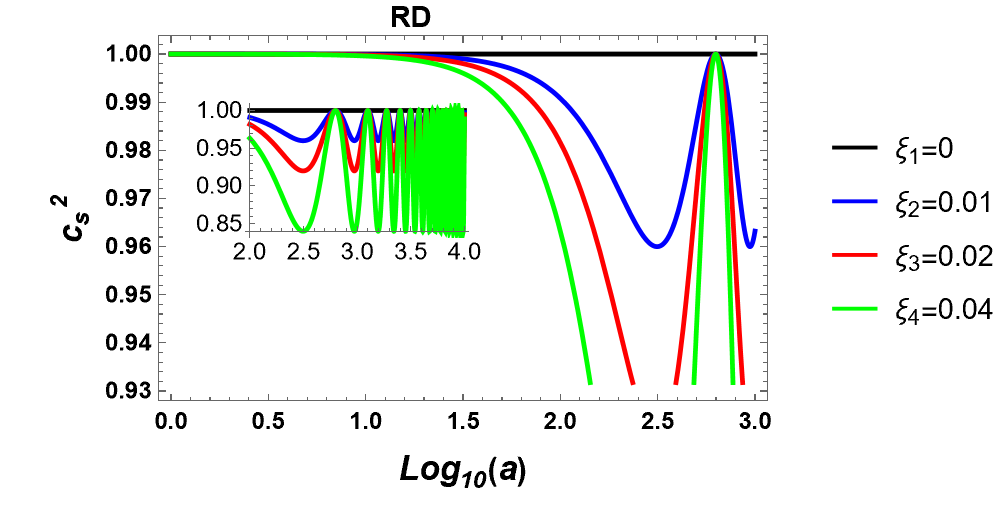}\\
	\qquad
	\includegraphics[width=0.45\linewidth]{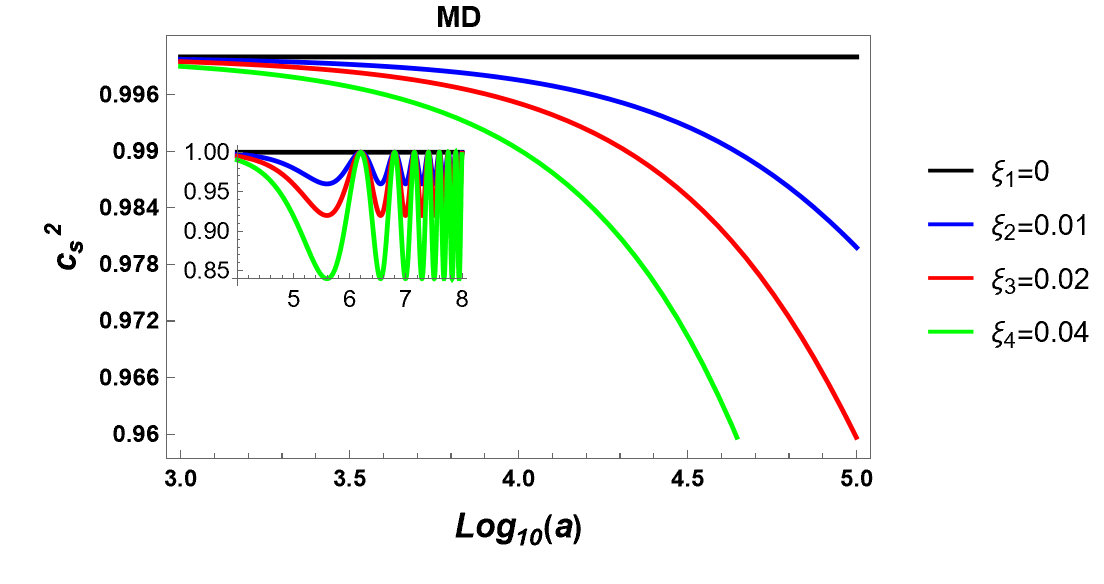}
	\qquad
	\caption{The plot of $c_S^2$ in different cosmological epochs with different $\xi$ parameters. The horizontal axis is $\log_{10}{(a)}$, and the vertical axis is $c_S^2$ . The black ($\xi=0$), blue ($\xi=0.01$), red ($\xi=0.02$), and green ($\xi=0.04$) lines represent various evolution of $c_S^2$. We also have set $k=1$ during inflation, $k=0.01$ during the RD epoch, and $k=0.005$ during the MD epoch.}
	\label{fig: c_s^2}
\end{figure}

Here, we list the formulas of non-trivial sound speed during different epochs as follows, 

\begin{equation}
\begin{cases}
\mbox{Inflation:}&a(\eta)=-\eta^{-1}\\&c_S^2=1-2\xi[1-\cos{(k10^{-y})}] \\ \mbox{RD:}&a(\eta)=\eta\\&c_S^2=1-2\xi[1-\cos{(k10^{y})}]\\ \mbox{MD:}&a(\eta)=\eta^2\\&c_S^2=1-2\xi[1-\cos{(k10^{\frac{y}{2}})}]
\end{cases}
\end{equation}
where we have used the $y=\log_{10}a$ and Eq. \eqref{formula of eta}. Once obtaining the formula of SSR in various epochs, we cloud also give their corresponding numeric given by Fig. \ref{fig: c_s^2}, where the values of $\xi$ cannot be large due to the observational constraints. The first panel of Fig. \ref{fig: c_s^2} clearly shows there is oscillation of $c_S^2$ with various values of $\xi$, in which the larger values of $\xi$, the more oscillation of $c_S^2$. As for the MD and RD, it will show the similar trend as shown in the second and the third panel of Fig. \ref{fig: c_s^2}.

\section{The Hamiltonian}
\label{the hamiltonian}
Before introducing the second order action of Mukhannov-Sasaki variable with the non-trivial sound speed, we need some concepts of curvature perturbation of inflaton. First, we need the perturbation of FLRW metric \eqref{bacground metric} as follows,
\begin{equation}
ds^2=a^2(\eta)\bigg[-(1+\psi(x_\mu))d\eta^2+(1-\psi(x_\mu))d\vec x^2\bigg]
\end{equation}
where $\psi(x_\mu)$ is the perturbation of FLRW metric. The second concept is the perturbation of inflaton 
\begin{equation}
\phi(x_\mu)=\bar{\phi}(\eta)+\delta\phi(x_\mu),
\label{perturbation of inflaton}
\end{equation}
where $\bar{\phi}(\eta)$ is the background field of inflaton and $\delta\phi(x_\mu)$ is the perturbation of inflaton. Being armed with metric \eqref{perturbation of inflaton}, we could obtain the action of the curvature perturbation as follows,
\begin{equation}
S=\frac{1}{2}\int dtd^3\vec x\frac{\dot\phi^2}{H^2}\bigg[\mathcal{R}-\frac{(\partial_i\mathcal{R})^2}{a^2}\bigg],
\label{action of curvature}
\end{equation}
where $H=\frac{\dot a}{a}$, $\mathcal{R}=\psi+\frac{H}{\dot\phi_0}\delta\phi$. This action can be transformed into the action in terms of Mukhannov-Sasaki variable $f$ and $c_S$. By the transformation $f=z\mathcal{R}$ and $z=\frac{\sqrt{2\epsilon}a}{c_S}$ ($\epsilon=-\frac{\dot H }{H^2}=1-\frac{\mathcal{H}'}{\mathcal{H}^2}$), then the action \eqref{action of curvature} will become as follows,
\begin{equation}
S=\frac{1}{2}\int d\eta d^3\vec{x}\bigg[f'^2-c_S^2(\partial_if)^2+\big(\frac{z'}{z}\big)^2f^2-2\frac{z'}{z}f'f-a^2V,_{\phi\phi}\bigg],
\label{total action}
\end{equation}
where prime denotes the derivative with respect to
the conformal time $\eta$ and $V,_{\phi\phi}=\frac{d^2V}{d\phi^2}$. Here, the action is valid for the single field inflation with non-trivial sound speed. The canonical momentum of action \eqref{total action} is defined as $\pi(\eta,\vec{x})=\frac{\delta \mathcal{L}}{\delta f'}$ and the Hamiltonian can be obtained 
\begin{equation}
H=\int d\eta d^3x(\pi f'- L)=\frac{1}{2}\int d\eta d^3x(\pi ^2 + c^2_S(\partial _if)^2+\frac{z'}{z}(\pi f+f\pi )+a^2V,_{\phi \phi}f^2), 
\label{total hamiltonian}
\end{equation}
where we have assumed that $\epsilon$ is a constant. Nextly, we could use the Fourier decomposition of operator as follows, 
\begin{equation}
\hat{f}(\eta,\vec{x})=\int{\frac{d^3k}{(2\pi)^{3/2}}}\sqrt{\frac{1}{2k}}\bigg[\hat{c}^{\dagger}_{\vec{k}}v^*_{\vec{k}}e^{-i\vec{k}\cdot\vec{x}}+\hat{c}_{\vec{k}}v_{\vec{k}}e^{i\vec{k}\cdot\vec{x}}\bigg]
\end{equation}

\begin{equation}
\hat{\pi}(\eta,\vec{x})=i\int{\frac{d^3k}{(2\pi)^{3/2}}}\sqrt{\frac{k}{2}}\bigg[\hat{c}^{\dagger}_{\vec{k}}u^*_{\vec{k}}e^{-i\vec{k}\cdot\vec{x}}-\hat{c}_{\vec{k}}u_{\vec{k}}e^{i\vec{k}\cdot\vec{x}}\bigg]
\end{equation}
where $\hat{c}_{-\vec{k}}^\dagger$ and $\hat{c}_{\vec{k}}$ represent the creation and annihilation operator. Making an appropriate normalization for $u_{\vec{k}}(\eta)$ and $v_{\vec{k}}(\eta)$, one can derive the following Hamiltonian 
\begin{equation}
\begin{split}
\hat{H_k}&=\bigg[\big(\frac{a^{2}V,_{\phi\phi}}{2k}+\frac{z'}{z}i-\frac{k}{2}+\frac{kc_S^2}{2}\big){\hat{c}^{\dagger} _{\vec{k}}}{\hat{c}^{\dagger} _{\vec{-k}}}+\big(\frac{a^2V,_{\phi\phi}}{2k}-\frac{z'}{z}i-\frac{k}{2}+\frac{kc_S^2}{2}\big){\hat{c} _{\vec{k}}}{\hat{c}_{\vec{-k}}}\\&+\big(\frac{a^2V,_{\phi\phi}}{2k}+\frac{k}{2}+\frac{kc_S^2}{2}\big)\big({\hat{c} _{\vec{k}}}{\hat{c}^{\dagger} _{\vec{k}}}+{\hat{c} _{-\vec{k}}}{\hat{c}^{\dagger} _{-\vec{k}}}\big)\bigg].\label{final hamiltonian}
\end{split}
\end{equation}
where we have transformed into the momentum space $k$ by the Fourier transformation. One can easily check our the Hamiltonian will nicely recover into Eq. (4.10) of Ref. \cite{Zhai:2024odw} as $c_S^2=1$.

\section{Lanczos algorithm for the open system}
\label{lanczos algorithm}
As the previous discussions, it is well known that our universe is an open system. In this section, we will follow the notation of Ref. \cite{Zhai:2024odw} to introduce the Lanczos algorithm for the open system, particularly, we will use the Arnoldi iteration via the Heisenberg form to introduce this algorithm \cite{Bhattacharya:2022gbz,Bhattacharjee:2022lzy,Liu:2022god,Nizami:2023dkf,Nandy:2024evd}. First, we should introduce the general Lindblad master equation as follows,
\begin{equation}
\dot\rho=-i[H,\rho]+\sum_{k}[L_k\rho L_k^\dagger-\frac{1}{2}\{L_k^\dagger L_k,\rho\}],
\label{general Lindblad}
\end{equation}
where $\rho$ is the density matrix, $H$ is the Hamiltonian, and $L_k^\dagger$ denotes the jump operator corresponding to the creation operator in this work. Since the inflationary universe is of very high temperature, we will focus on the an infinite temperature environment, characterized by the maximally mixed state. Under this approximation, we will have 
\begin{equation}
\sum_{k}[L_k\rho_\infty L_k^\dagger-\frac{1}{2}\{L_k^\dagger L_k,\rho_\infty\}]=0,
\label{assumption for rho}
\end{equation} 
where $\rho_\infty$ denotes the infinite temperature-state. Consequently, the operator dynamics is governed by 
\begin{equation}
\mathcal{\hat O}(t)=\exp(i\mathcal{L}t)\mathcal{\hat O},
\label{dynamic of operator}
\end{equation}
where  the Heisenberg-picture Lindbladian $\mathcal{L}$ is given by \cite{Bhattacharya:2022gbz}, 
\begin{equation}
\mathcal{L}=\mathcal{L}_H+\mathcal{L}_D,~~\mathcal{L}_H\mathcal{\hat O}=[H,\mathcal{\hat O}],~~\mathcal{L}_D=	\sum_{k}[L_k\mathcal{\hat O} L_k^\dagger-\frac{1}{2}\{L_k^\dagger L_k,\mathcal{\hat O}\}]
\end{equation}
where $\mathcal{L}_H$ response the Hermitian dynamics, $\mathcal{L}_D$ denotes the dissipative part, and $\mathcal{L}$ bahaves as a superoperator on the operator space. Furthermore, we could define inner product of operator that is naturally related to the generalized Kyrlov basis, namely $(A|B) := \mathrm{Tr}[\rho_\infty A^\dagger B]$. 

Once we have the dynamics of operator, we further need the appropriate basis for calculating this kind of physical observable. Following \cite{Bhattacharya:2022gbz}, we will implement the orthonormal basis $\{\mathcal{O}_0, \ldots, \mathcal{O}_n, \ldots\}$, which is generated by the open-system Lindbladian according to the following procedure:
\begin{equation}
\rm span(\mathcal{O}_0,...,\mathcal{O}_n)=span (\mathcal{V}_0,...,\mathcal{V}_n), 
\label{span of krylov basis}
\end{equation} 
where $\mathcal{V}_n$ is the original Krylov basis. This generalized Krylov basis will be obtained by the so-called Arnoldi iteration. This procedure is started by an initial vector $\mathcal{O}_0 \propto \mathcal{O}$. Thereafter, we could construct the the vector within this basis, 
\begin{equation}
|\mathcal{U}_k)=\mathcal{L}| \mathcal{O}_{k-1}).
\label{recursion relation}
\end{equation}
Then, the iteration is performed as follows: $1.~ h_{j,k-1}=(\mathcal{O}_j|\mathcal{U}_k)$, $2.~|\tilde{\mathcal{U}}_k)=|\mathcal{U}_k)-\sum_{j=0}^{k-1}h_{j,k-1}|\mathcal{O}_j)$, $3.~h_{k,k-1}=\sqrt{(\tilde{U}_k | \tilde{\mathcal{U}}_k)}$, it will stop as if $h_{k,k-1}=0$, otherwise, we could define
\begin{equation}
|\mathcal{O}_k)=\frac{|\tilde{\mathcal{U}}_k)}{h_{k,k-1}}.
\label{final recursion}
\end{equation}
Consequently, the Lindbladian will change into an upper Heisenberg form in the Krylov basis (also called the Arnoldi basis)
\begin{equation}
\mathcal{L}=
\begin{pmatrix}
h_{0,0} & h_{0,1} & h_{0,2} & \cdots & \cdots & h_{0,n} \\
h_{1,0} & h_{1,1} & h_{1,2} & \cdots & \cdots & h_{1,n} \\
0 & h_{2,1} & h_{2,2} & h_{2,3} & \cdots & \cdots \\
\vdots & 0 & h_{3,2} & \cdots & \cdots & \cdots \\
0 & \vdots & 0 & \vdots & \ddots & h_{n-1,n} \\
0 & 0 & \vdots & 0 & h_{n,n-1} & h_{n,n}
\end{pmatrix}
\label{matrix of L}
\end{equation}
where $h_{m,n}=(\mathcal{O}_m|\mathcal{L}|\mathcal{O}_n)$. As $\mathcal{L}$ is a Hermitian superoperator, the Arnoldi iteration will reduce into the Lanczos algorithm. In this work, we will focus on the following Liouvillian super-operator which could describe the open system dubbed as the Lindbladian $\mathcal{L}_{o}$,
\begin{equation}
\mathcal{L}_{o}|\mathcal{O}_{n})=-ic_{n}|\mathcal{O}_{n})+b_{n+1}|\mathcal{O}_{n+1})+b_{n}|\mathcal{O}_{n-1})
\label{eq Lindbladian}
\end{equation}
where the coefficient $c_n$ encodes information about the open system corresponding to the $h_{n,n}$ of matrix \eqref{matrix of L}, the Lanczos coefficient $b_n$ corresponds to $h_{n,n-1}$ of \eqref{matrix of L}, and the Lindbladian $\mathcal{L}_{o}$ is equivalent to the Hamiltonian when represented by the creation and annihilation operators \cite{Parker:2018yvk}. Thereafter, one could express the operator as follows,
\begin{equation}
\mathcal{O}(t)=e^{i\mathcal{L}t}\mathcal{O}=\sum_{n=0}^{\infty} (i)^{n}\phi_{n}(t)|\mathcal{O}_{n}),
\label{O with O_n}
\end{equation}
where $\phi_n$ is the wave function satisfied with $\Sigma_{n}|\phi|^2=1$. Combine Eq. \eqref{O with O_n} and Schr$\Ddot{o}$dinger equation, we could derive the following equation, 
\begin{equation}
\partial_\eta\phi+2b_n\partial_n\phi+\tilde{c}_n\phi=0,
\label{equ of cn}
\end{equation}
where we have defined \(\widetilde{c}_n = -i c_n\). In the continuous limit, we employ the approximations \(b_{n+1} \approx b_n\) and \(\phi_{n+1} - \phi_{n-1} \approx 2 \partial_x \phi\).  Next, we could define the Krylov complexity and Krylov entropy according to $\phi_n$,
\begin{equation}
C_K=\sum_{n} n|\phi_{n}|^{2},~~~K_E=-\Sigma_{n=0}^{+\infty}|\phi_n|^2\ln|\phi_n|^2,
\label{krylov complexity and krylov entropy}
\end{equation}
where $C_K$ denotes the Krylov complexity and $K_E$ represents the Krylov entropy. Due to the additional coefficient $c_n$, the Hamiltonian \eqref{final hamiltonian} can be divided into two parts 
\begin{equation}
\mathcal{H}_O=\mathcal{H}_{closed}+\mathcal{H}_{open},
\label{speration of hamiltonian}
\end{equation}
In light of \eqref{eq Lindbladian}, we can easily obtain the open part (generating $c_n$) of Hamiltonian, 
\begin{equation}
\mathcal{H}_{open}=
(k+\frac{a^{2}V_{,\phi\phi}}{2k})(\hat{c}_{-\vec{k}}^{\dagger}\hat{c}_{-\vec{k}}+\hat{c}_{\vec{k}}\hat{c}_{\vec{k}}^{\dagger})
\label{eq H open}
\end{equation}
and the close part (generating $b_n$ and $b_{n-1}$) is
\begin{equation}
\mathcal{H}_{close}=
(\frac{a^{2}V_{,\phi\phi}}{2k}+i\frac{z'}{z})\hat{c}_{\vec{k}}^{\dagger}\hat{c}_{-\vec{k}}^{\dagger}+(\frac{a^{2}V_{,\phi\phi}}{2k}-i\frac{z'}{z})\hat{c}_{\vec{k}}\hat{c}_{-\vec{k}}
\label{eq H close}
\end{equation}
where $|\mathcal{O}_{n})=\Bigl | n_{\vec{k}};n_{-\vec{k}}\Bigr \rangle$. By plugging Eq. \eqref{eq H open} and Eq. \eqref{eq H close} into Eq. \eqref{eq Lindbladian}, it is easy to obtain for $c_{n}$ and $b_{n}$ as
\begin{equation}
c_{n}=
i\bigg(\frac{a^2V,_{\phi\phi}}{2k}+\frac{k}{2}+\frac{kc_S^2}{2}\bigg)(2n+1),
\label{eq cn}
\end{equation}
\begin{equation}
b_{n}=
n\sqrt{\bigg(\frac{a^2V_{\phi\phi}}{2k}-\frac{k}{2}+\frac{kc_S^2}{2}\bigg)^2+\big(\frac{z'}{z}\big)^2},
\label{eq bn1}
\end{equation}
where $n$ is the quantum number. Following Ref. \cite{Zhai:2024odw}, one observes that $c_n$ is strongly associated with dissipative effects in dynamical systems, which will be explicitly related to the dissipative coefficient. 

\subsection{Statistical property of universe}
\label{statistical property of universe}

The coefficient $b_n$ characterizes chaotic behavior, with the specific form $b_n = \alpha n + \gamma$ (where $\alpha$ and $\gamma$ are Hamiltonian-dependent parameters) indicating an infinite, non-integrable, many-body system \cite{Parker:2018yvk}. Based on \eqref{eq bn1}, we conclude that our universe is an infinite, non-integrable, many-body system, where
\begin{equation}
\alpha = \sqrt{
	\left( \frac{a^2 V_{\phi\phi}}{2k} - \frac{k}{2} + \frac{k c_S^2}{2} \right)^{\!\!2}
	+ \left( \frac{z'}{z} \right)^{\!2}
} \quad \text{and} \quad \gamma = 0.
\end{equation}
To better describe the statistical properties of the universe, we follow the notation of \cite{Zhai:2024odw,Parker:2018yvk,Li:2024kfm,Li:2024iji,Li:2024ljz} and define:
\begin{equation}
b_{n}^{2} = \lvert 1 - u_{1}^{2} \rvert \, n(n - 1 + \beta), \quad 
c_{n} = i u_{2} (2n + \beta)
\label{eq bn and cn}
\end{equation}
where $u_2$ denotes the dissipative coefficient. Relating to \eqref{eq cn} and \eqref{eq bn1}, we obtain:
\begin{equation}
\begin{cases}
\displaystyle
\lvert 1 - u^2_1 \rvert = 
\left( \frac{a^2 V,_{\phi\phi}}{2k} - \frac{k}{2} + \frac{k c_S^2}{2} \right)^{\!\!2}
+ \left( \frac{z'}{z} \right)^{\!2}
\\[2.5ex]
\displaystyle
u_2 = \frac{a^2 V,_{\phi\phi}}{2k} + \frac{k}{2} + \frac{k c_S^2}{2}
\\[2.5ex]
\displaystyle
u_2' = \frac{a a' V,_{\phi\phi}}{k} + \frac{a^2 V,_{\phi\phi}'}{2k} + k c_S c_S'
\end{cases}
\label{eq mu1 mu2}
\end{equation}
with $\beta = 1$. The coefficient $u_2$ represents the dissipative property of the dynamical system:
\begin{itemize}
	\item $u_2 < 1$: weak dissipation
	\item $u_2 \geq 1$: strong dissipation
\end{itemize}
Meanwhile, $b_n$ is equivalent to the Lyapunov exponent $\lambda$ through $\lambda = 2\alpha$. For $n=2$, this corresponds to $\lambda = b_2$ with $\alpha$ given above. Using \eqref{eq mu1 mu2}, we compute the numerical values of $b_2 = \lambda$ and $u_2$.

\begin{figure}
	\centering
	\includegraphics[width=.4\textwidth]{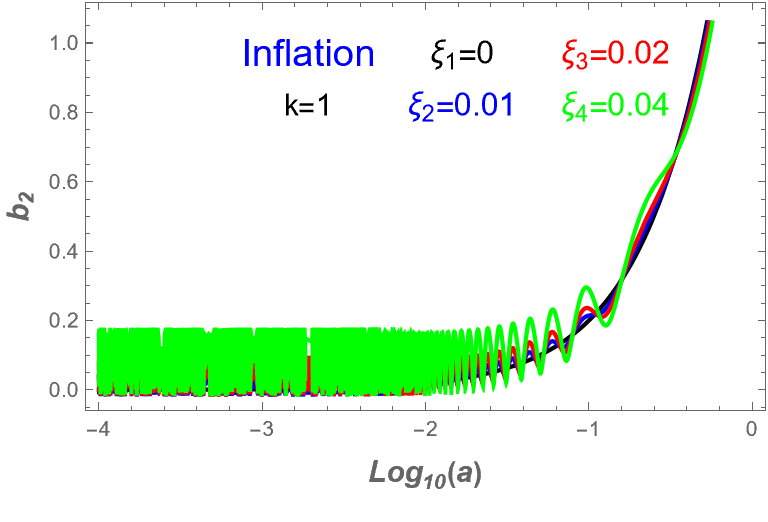}\\
	\qquad
	\includegraphics[width=.4\textwidth]{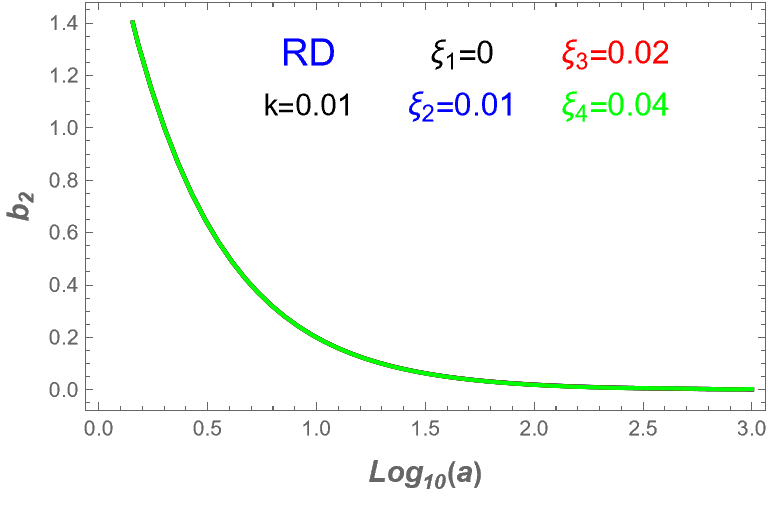}
	\qquad
	\includegraphics[width=.4\textwidth]{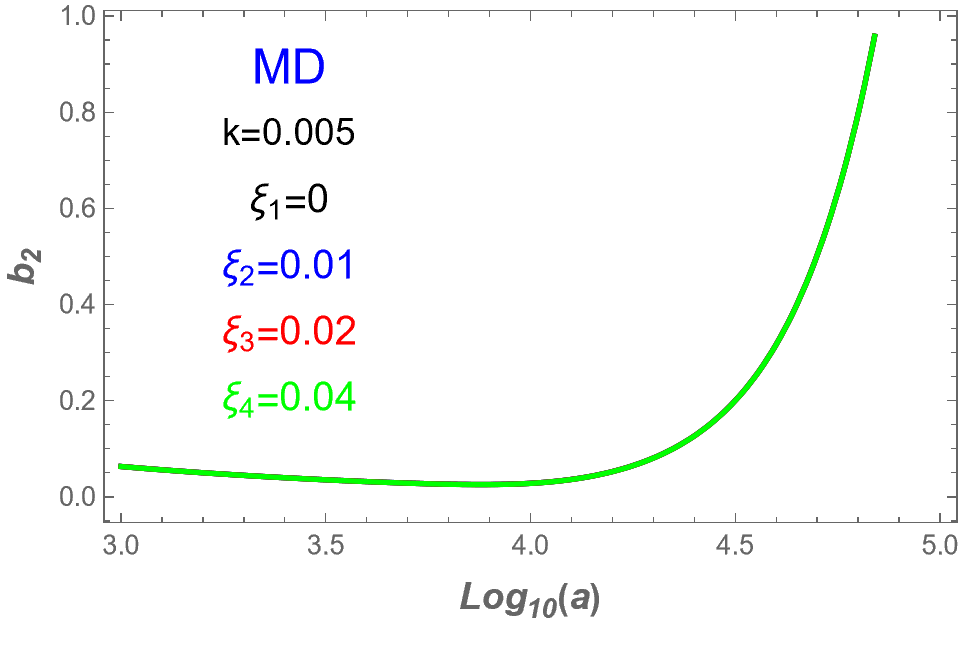}
	\caption{The plots $b_2$ against $\log_{10}a$ for $n=2$, showing numerical values across three cosmological epochs: inflation, RD, and MD. For simplicity, we adopt $H_0=1$ and set the  $k=1$ during inflation, 
		$k=0.01$ during RD, and $k=0.005$ during MD.}
	\label{fig: op s bn}
\end{figure}

Figure~\ref{fig: op s bn} clearly shows that $b_2 = \lambda$ (representing the universe's chaotic feature) is significantly enhanced during inflation due to exponential expansion. This indicates dramatically increased chaos in the inflationary epoch, where higher $\xi$ values yield more pronounced oscillations. In contrast, both RD and MD eras exhibit nearly identical $b_2$ trends across different $\xi$ values: $b_2$ decreases during RD but increases again during MD. The overall behavior of $b_2$ is consistent with previous studies~\cite{Zhai:2024odw,Li:2024kfm}.

\begin{figure}
	\centering
	\includegraphics[width=.5\textwidth]{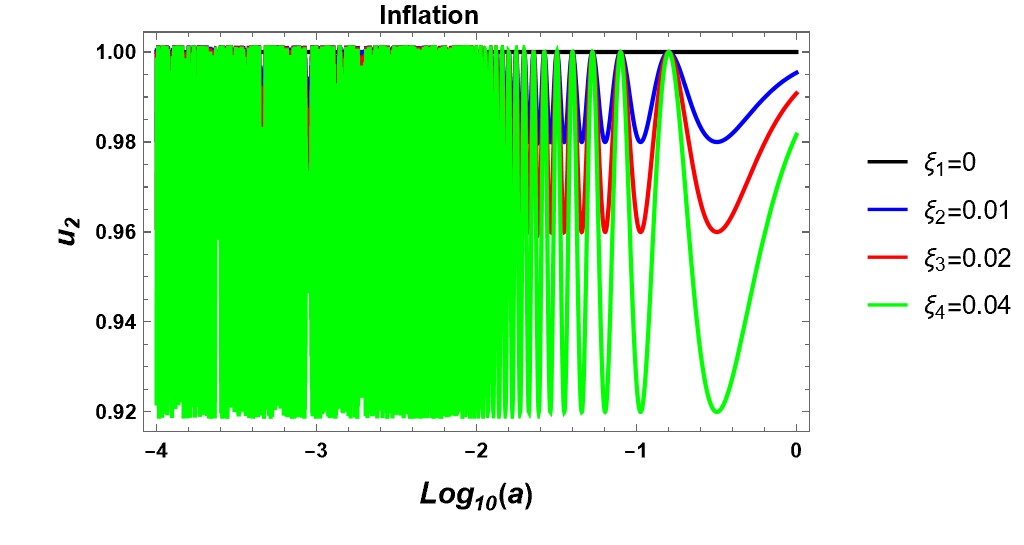}
	\qquad
	\includegraphics[width=.45\textwidth]{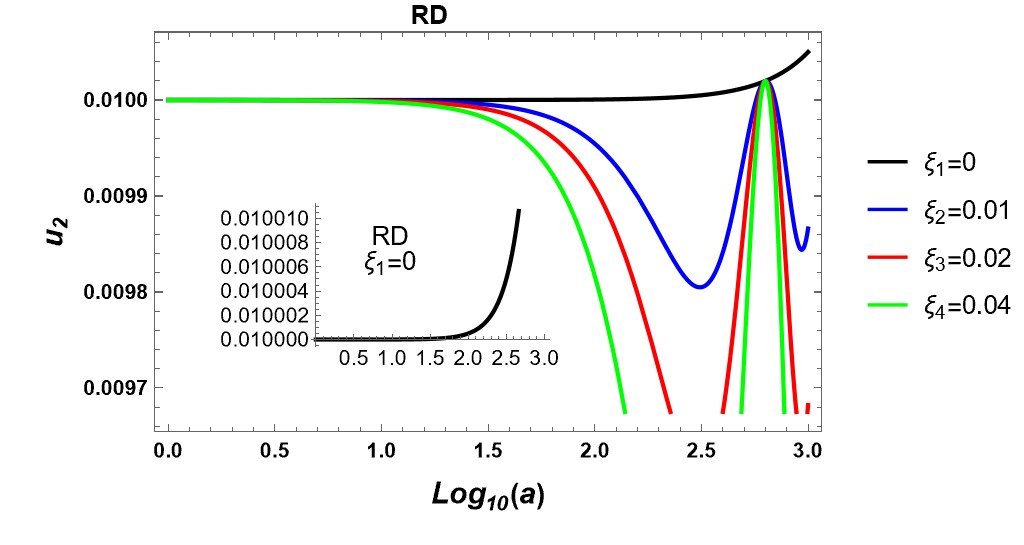}
	\qquad
	\includegraphics[width=.45\textwidth]{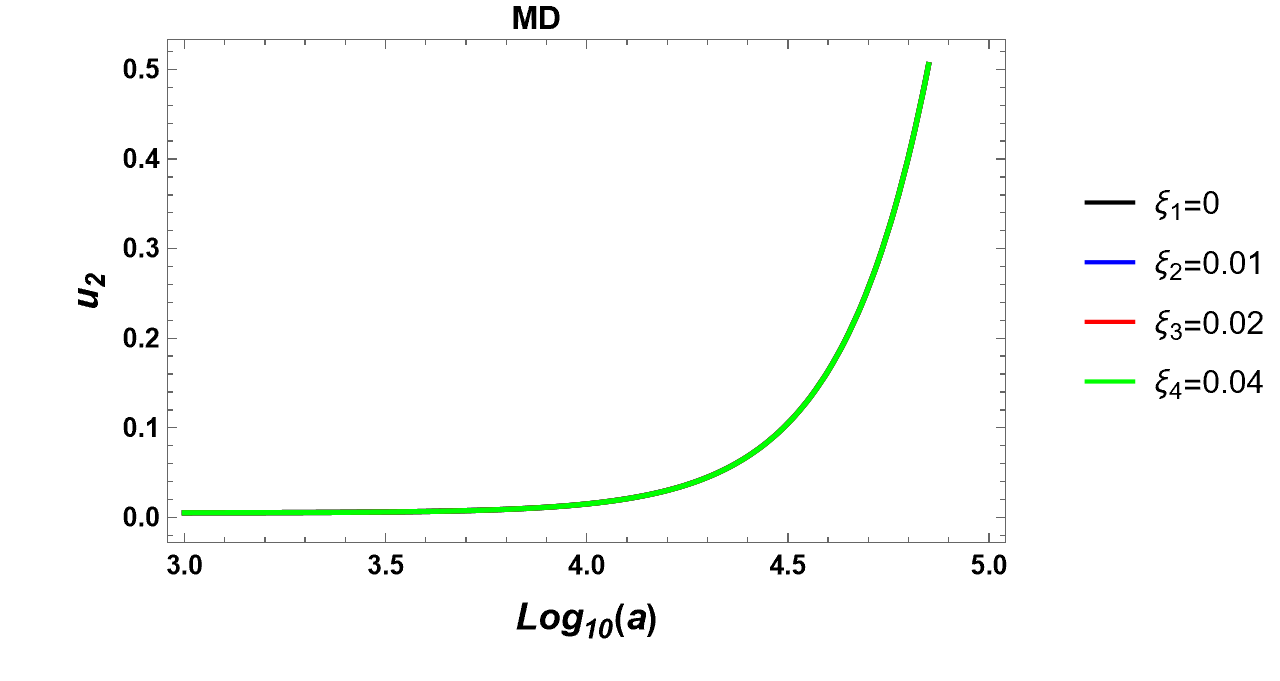}
	\caption{ The numerical results for the dissipation coefficient $u_2$ as a function of $\log_{10}a$ a during the RD and MD epochs. For simplicity, we adopt $H_0=1$ and set the $k=1$ during inflation, $k=0.01$ during RD, and $k=0.005$ during MD.}
	\label{fig: op s u2}
\end{figure}
Figure~\ref{fig: op s u2} illustrates the evolution of $u_2$ during inflation, RD, and MD epochs. During inflation, the system exhibits strong dissipation, consistent with previous studies~\cite{Zhai:2024odw,Parker:2018yvk,Li:2024kfm,Li:2024iji,Li:2024ljz}. Post-inflation, the universe transitions to a weak dissipative phase. While non-trivial sound speed introduces oscillations in the evolution of $u_2$, it does not alter the fundamental nature of $u_2$ in any epoch.

\section{Krylov complexity and Krylov entropy }
\label{krylov complexity and krylov entropy}
In this section, we employ the open system method \cite{Zhai:2024odw} to study the Krylov complexity and Krylov entropy in single-field inflation with non-trivial sound speed. The essential element for obtaining the corresponding numerical results is the evolution of the two squeezing parameters, $r_k(\eta)$ and $\phi_k(\eta)$, which are determined by the specific Hamiltonian \eqref{final hamiltonian} and the open two-mode squeezed state (OTMSS). Having derived the target Hamiltonian \eqref{final hamiltonian}, we proceed to introduce the OTMSS in the following subsection.

\subsection{The wave function for the open system}
\label{the wave function for the open system}
As Ref. \cite{Grishchuk:1990bj} discovered, the curvature perturbation from the quantum level to the classical level is a natural squeezing process. Thus, the two-mode squeezed state is perfect candidate to simulate the evolution of quantum process. However, we will show the OTMSS is more generic and more complete compared to the two-mode squeezed state. First and foremost, the formula of OTMSS is presented by 
\begin{equation}
|\mathcal{O}(\eta))=\frac{ \rm sech r_k(\eta)}{1+u_2\tanh r_k(\eta)}\sum_{n=0}^{\infty}|1-u_1^2|^{\frac{n}{2}}\frac{(-\exp(2i\phi_k(\eta))\tanh r_k(\eta))^{n}}{(1+u_2\tanh r_k( \eta))^{n}}|n,n\rangle_{-\vec{k},\vec k},
\label{wave function1}
\end{equation}
where $|n,n\rangle_{-\vec k,\vec k}$ denotes the Arnoldi basis used in this work, $\eta$ is the conformal time, $r_k$ and $\phi_k$ are the squeezing parameters, and $u_2$ is the dissipative coefficient, where $u_2$ and $|1-u_1^2|$ are determined by the Hamiltonian, their formula is from \eqref{eq mu1 mu2}. The detailed calculation of OTMSS can be found in \cite{Zhai:2024odw}. By taking place of two-mode squeezed state, we will utilize the OTMSS to represent the quantum fluctuations in the entire early period, including inflation, RD, MD. The OTMSS \eqref{wave function1} is a wave function that could describe the dissipative effects of open system, and it will reduce to the two-mode squeezed state in the weak dissipative approximation $(u_2\ll 1)$,
\begin{equation}
|\mathcal{O}(\eta))= \frac{1}{\cosh r_k}\sum_{n=0}^{\infty} (-1)^ne^{2in\phi_k}\tanh^nr_k|n;n\rangle_{\vec{k},-\vec{k}}+\mathcal{O}^n(u_2). 
\label{two mode squeezed state1}
\end{equation}
During the RD and MD eras, the dissipation coefficient $u_2$ remains significantly small ($u_2 \ll 1$), placing the universe in a weak dissipative phase. Consequently, while the OTMSS captures essential non-unitary information, the results in these later epochs do not deviate drastically from the 'closed system' approximation, as the open-system corrections remain perturbative. Here, we could see that the OTMSS is more genric and more complete compared to two-mode squeezed state, which incorporates the information of open system part for the Hamiltonian since its information is encoded in $u_2$. And the derivation of OTMSS is only related to the second kind Meixner polynomial \cite{Hetyei_2009}, whose formula is
\begin{equation}
P_{n+1}(x)=(x-\tilde{c}_n)P_n(x)-b_n^2P_{n-1}(x),
\label{eq definition Pn}
\end{equation}
where we define $\tilde{c}_n=-ic_n$, and $x$ represents the Hamiltonian. With the initial condition $P_0(x) =1$ and $P_1(x) = x-c_0$, and the polynomial $P_n(x)=\det (x-\mathcal{L}_n)$ corresponds to an open system, and $\mathcal{L}_n$ represents the Liouvillian superoperator for the $n$-th quantum state. Thereafter, we could introduce the basis equavallent to the Anorldi basis $\{e_n\}$, then we could represent $ |P_{n}(x))=\Bigl (\prod_{k=1}^{n}b_{k}\Bigr )|\mathcal{O}_n),\ \ \mbox{and}\ \ \ |x^{n})=\mathcal{L}^{n}|\mathcal{O})$. Combine with Eq. \eqref{eq definition Pn}, it can be derived further for Eq. \eqref{eq definition Pn},
\begin{equation}
b_{n+1}e_{n+1}+b_ne_{n-1}=(x-\tilde{c}_n)e_n,
\label{equalled relation of bn}
\end{equation}
where \( e_n = \mathcal{O}_n \) in our construction, which is equivalent to the generalized Lanczos algorithm given in Eq.~\eqref{eq Lindbladian}. Being armed with Eq. \eqref{equalled relation of bn}, we could construct the OTMSS beyond the pure two-mode squeezed state \eqref{two mode squeezed state1}. The conceptual advantage of OTMSS is including the dissipative effects of the specific Hamiltonian.

\subsection{Evolution of $r_k$ and $\phi_k$}
The Krylov complexity and Krylov entropy are depending on the $\phi_k$ and $r_k$, thus we need their corresponding evolution determined by the Schr\"{o}dinger equation
\begin{equation}
\hat{H} |\mathcal{O}(\eta)\rangle = i \partial_\eta |\mathcal{O}(\eta)\rangle.
\label{schrodinger equation}
\end{equation}
Based on it, we could see that the various Hamiltonian will yield the distinctive evolution equation of $r_k$ and $\phi_k$. Here, we only list their resulting evolution equations as follows,
\begin{equation}
\begin{split}
r'_k=\frac{|1-u_1^2|^{\frac{1}{2}}\sinh{2r_k}[(\frac{a^2V_{\phi\phi}}{2k}-\frac{k}{2}+\frac{kc_S^2}{2})\sin{2\phi_k}-\frac{z'}{z}\cos{2\phi_k}]-u'_2\sinh{2r_k}}{\sin{2r_k}+2u_2\cosh^2{r_k}}  
\end{split}
\label{rk evolution}
\end{equation}
\begin{equation}
\begin{split}
\phi'_k=&-\frac{1}{2}(k+kc^2_S+\frac{a^2V_{\phi\phi}}{k})+\frac{1}{2}[(\frac{a^2V_{\phi\phi}}{2k}-\frac{k}{2}+\frac{kc^2_S}{2})\cos{2\phi_k}+\frac{z'}{z}\sin{2\phi_k}]\\&(|1-u_1^2|^{-\frac{1}{2}}(u_2+\coth{r_k})+|1-u_1^2|^{\frac{1}{2}}\frac{\tanh{r_k}}{1+u_2\tanh{r_k}})
\end{split}
\label{phik evolution}
\end{equation}
where the prime denotes differentiation with respect to conformal time $\eta$. We emphasize that this is the first-time derivation for $\phi_k$ and $r_k$ under the non-trivial sound speed. The detailed calculations for this derivation can be found in Appendix \ref{Appendix A}. One can easily verify that the evolution equations for $r_k$ and $\phi_k$ reduce to the case studied in \cite{Zhai:2024odw} when $c_S = 1$ and $\epsilon$ is constant. From Eqs. \eqref{rk evolution} and \eqref{phik evolution}, it is evident that these equations describe the entire early universe, as they incorporate the influence of the inflationary potential and retain the Mukhanov-Sasaki variable. 

\begin{figure}
	\centering
	\includegraphics[width=.4\textwidth]{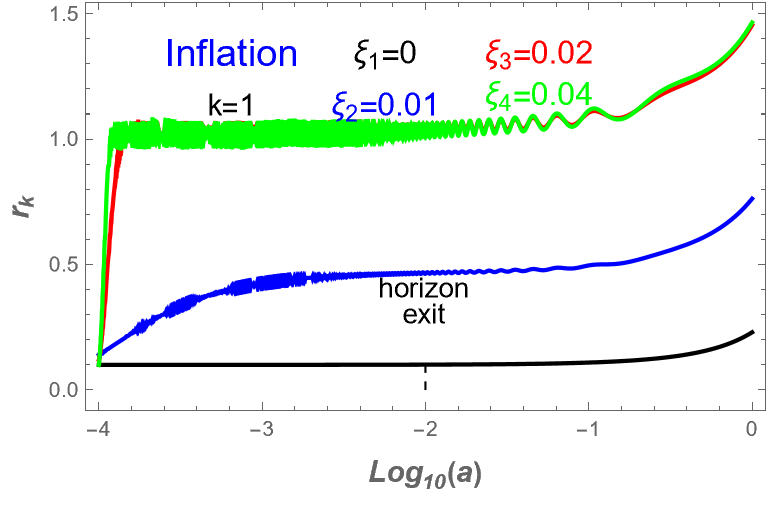}\\
	\qquad
	\includegraphics[width=.4\textwidth]{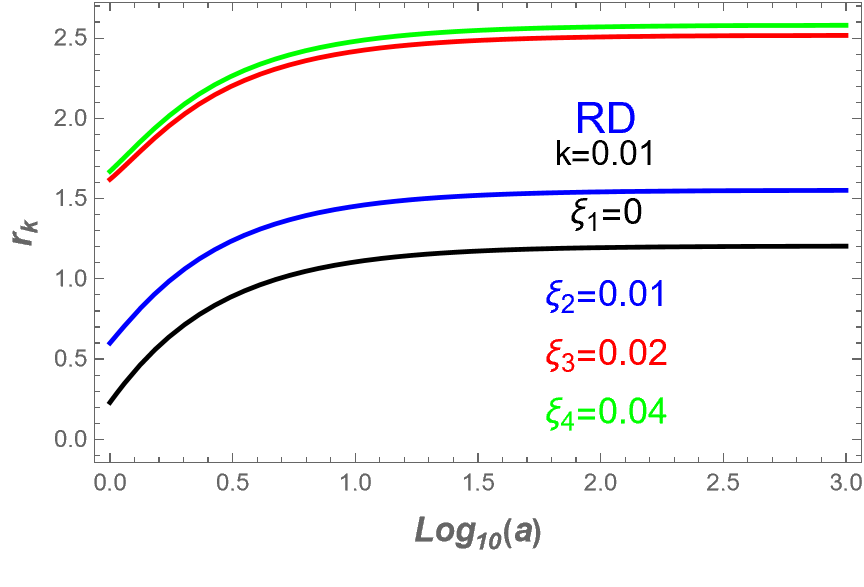}
	\qquad
	\includegraphics[width=.4\textwidth]{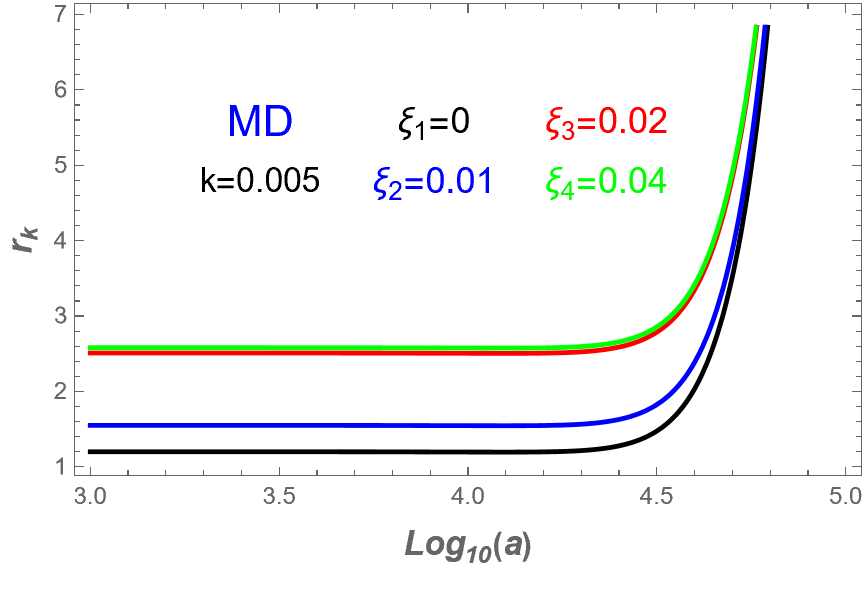}
	\caption{The numerical of $r_{k}$ in terms of $\log_{10}a$ for three different periods (inflation, RD, and MD), where we set $H_{0}=1$, $k=1$ in inflation, $k=0.01$ at RD and $k=0.005$ at MD for simplicity.}
	\label{fig: op s rk}
\end{figure}

Once obtaining the evolution equation of $\phi_k$ and $r_k$, we could further transform into the variable in terms of $y=\log_{10}a$, where $a$ is the scale factor, since it is more convenient compared with the conformal time explicitly. Here, we list the formula as follows,
\begin{equation}
\begin{split}
\frac{H_{0}}{\ln 10}10^{-\frac{y}{2}(1+3\omega)}\frac{dr_{k}}{dy}=&\frac{|1-u_1^2|^{\frac{1}{2}}\sinh{2r_k}(\frac{10^{2y}V_{\phi\phi}}{2k}-\frac{k}{2}+\frac{kc_S^2}{2})\sin{2\phi_k}}{\sin{2r_k}+2u_2\cosh^2{r_k}}\\&-\frac{|1-u_1^2|^{\frac{1}{2}}\sinh{2r_k}(H_010^{-\frac{y}{2}(1+3\omega)}-\frac{\xi k \sin{k\eta}}{1-2\xi[1-\cos{k\eta}]})\cos{2\phi_k}}{\sin{2r_k}+2u_2\cosh^2{r_k}}\\&-\frac{u'_2\sinh{2r_k}}{\sin{2r_k}+2u_2\cosh^2{r_k}}
\end{split}
\label{rk1}
\end{equation}
\begin{equation}
\begin{split}
\frac{H_{0}}{\ln 10}10^{-\frac{y}{2}(1+3\omega)}\frac{d\phi_{k}}{dy}=&-\frac{1}{2}(k+kc^2_S+\frac{10^{2y}V_{\phi\phi}}{k})+\frac{1}{2}[(\frac{a^2V_{\phi\phi}}{2k}-\frac{k}{2}+\frac{kc^2_S}{2})\cos{2\phi_k}\\&+(H_010^{-\frac{y}{2}(1+3\omega)}-\frac{\xi k \sin{k\eta}}{1-2\xi[1-\cos{k\eta}]})\sin{2\phi_k}]\\&(|1-u_1^2|^{-\frac{1}{2}}(u_2+\coth{r_k})+|1-u_1^2|^{\frac{1}{2}}\frac{\tanh{r_k}}{1+u_2\tanh{r_k}})
\end{split}
\label{phik1}
\end{equation}
where 
\begin{equation*}
\begin{split}
\begin{cases}
& u_{2}=\frac{k}{2}+\frac{kc_S^2}{2}+\frac{10^{2y}V_{,\phi\phi}}{2k},\\&  u_{2}'=H_{0}10^{-\frac{y}{2}(1+3\omega)}\frac{10^{2y}V_{,\phi\phi}}{k}+\frac{10^{2y}V_{,\phi\phi}'}{2k}+kc_Sc_S'\\ 
&\ |1-u_{1}^{2}|=(\frac{10^{2y}V_{,\phi\phi}}{2k}-\frac{k}{2}+\frac{kc_S^2}{2})^{2}+(H_{0}10^{-\frac{y}{2}(1+3\omega)}-\frac{\xi k \sin{k\eta}}{1-2\xi[1-\cos{k\eta}]})^{2}. 
\end{cases}
\end{split}
\end{equation*}
Then, we could give their numeric according to Eqs. \eqref{rk1} and \eqref{phik1}. Fig. \ref{fig: op s rk} clearly indicates that evolution trend of $r_k$. Here, we will give the its corresponding evolution in three periods. 

\textbf{Inflation}: The overall trend of $r_k$ is exponential growth, consistent with previous findings for the case $\xi = 0$, which corresponds to single-field inflation \cite{Zhai:2024odw}. Increasing $\xi$ enhances the amplitude of $r_k$. The enhancement of the squeezing amplitude $r_k$ in the presence of a non-trivial sound speed can be understood through the modified kinetic structure in Eq. \eqref{ssr}. When $c_S^2$ deviates from unity, it effectively modulates the frequency of the Mukhannov-Sasaki variable. This modulation triggers a parametric resonance-like effect during the inflationary expansion, leading to the enhancement into the squeezing of quantum states compared to the standard canonical case ($c_S = 1$). During the \textbf{RD} era, $r_k$ exhibits similar behavior across different values of $\xi$: it grows and eventually saturates at a constant value. Higher values of $\xi$ further amplify the saturation amplitude of $r_k$. In the \textbf{MD} era, the amplitude of $r_k$ is also enhanced with larger $\xi$. The overall behavior remains consistent: $r_k$ reaches a nearly constant value before resuming growth for $y > 4.5$. 

\begin{figure}
	\centering
	\includegraphics[width=.4\textwidth]{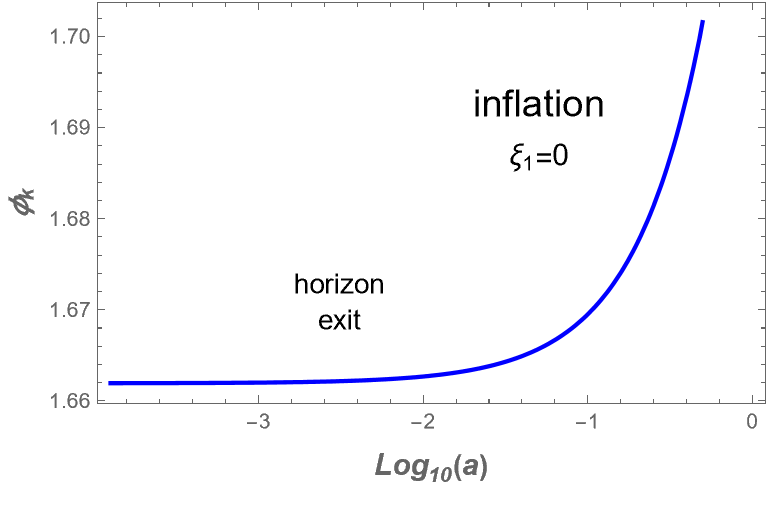}
	\qquad
	\includegraphics[width=.4\textwidth]{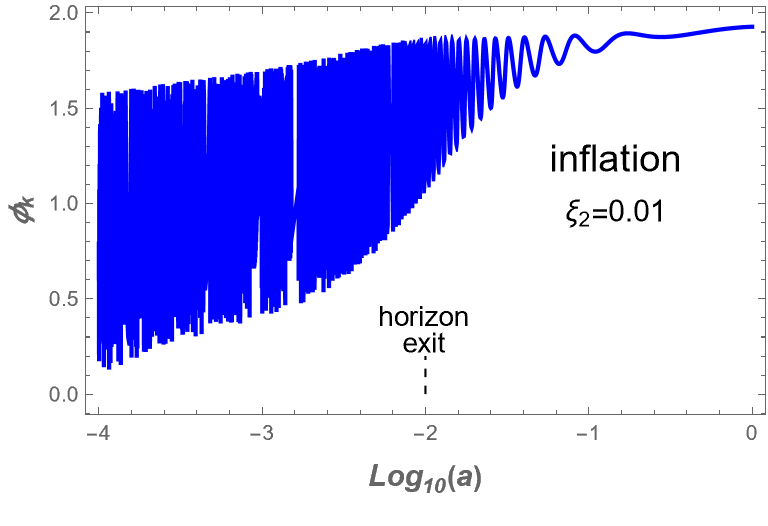}\\
	\qquad
	\includegraphics[width=.4\textwidth]{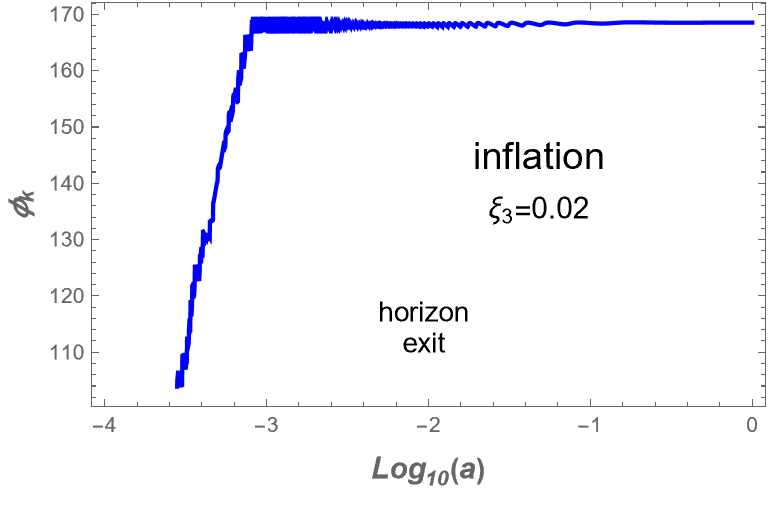}
	\qquad
	\includegraphics[width=.4\textwidth]{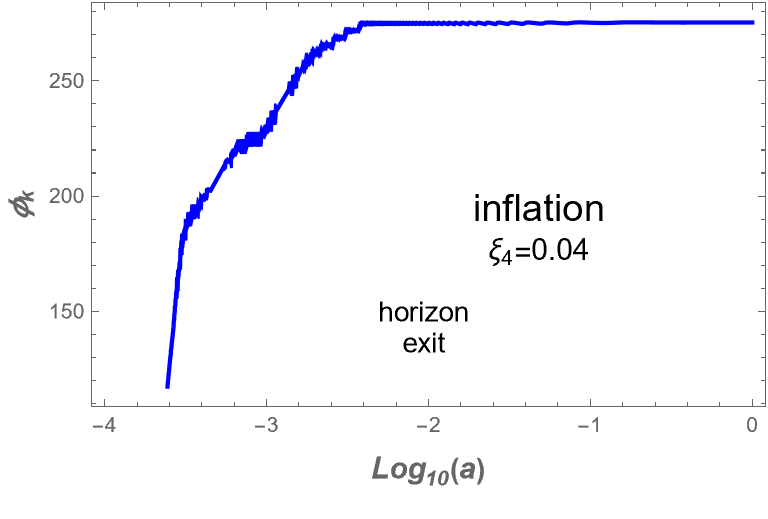}
	\caption{The numerical of $\phi_{k}$ in terms of $\log_{10}a$ for  inflation period , where we set $H_{0}=1$, $k=1$  for simplicity.}
	\label{fig: op s phi}
\end{figure}

\begin{figure}
	\centering
	\includegraphics[width=.4\textwidth]{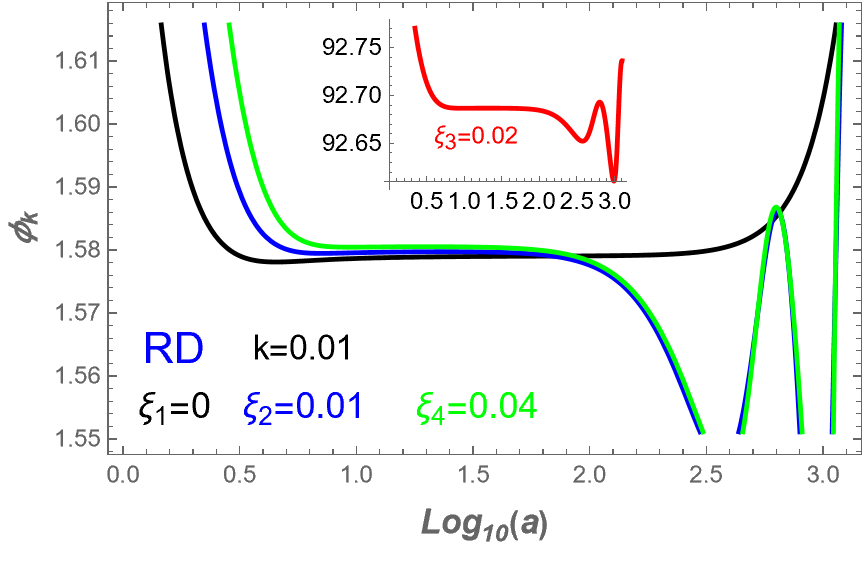}
	\qquad
	\includegraphics[width=.4\textwidth]{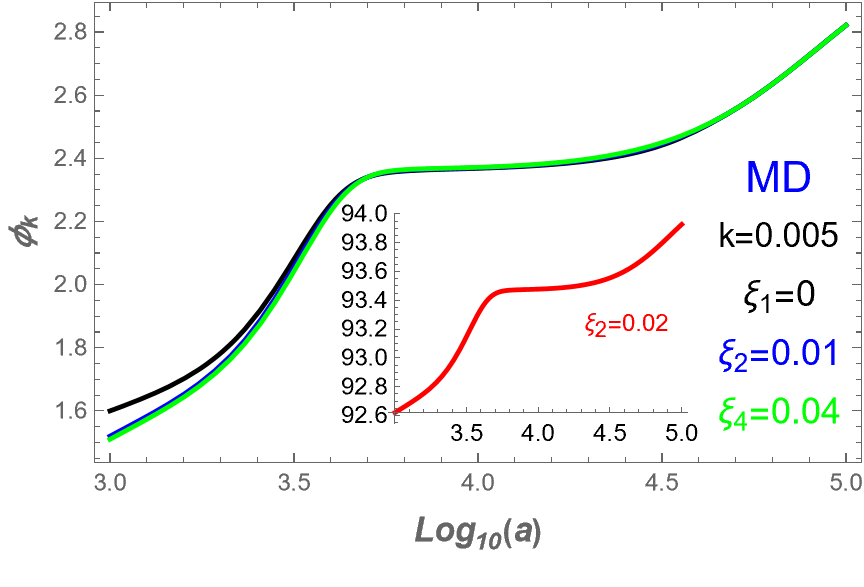}
	\caption{The numerical of $\phi_{k}$ in terms of $\log_{10}a$ for RD, and MD periods, where we set $k=0.01$ at RD and $k=0.005$ at MD for simplicity.}
	\label{fig: op s phi1}
\end{figure}

Regarding the evolution of $\phi_k$, Figs. \ref{fig: op s phi} and \ref{fig: op s phi1} clearly illustrate its behavior as a function of $y$, highlighting significant differences due to inflation. We also provide corresponding numerical values across various epochs. During \textbf{Inflation}, Fig. \ref{fig: op s phi} shows that $\phi_k$ undergoes substantial changes, with its amplitude being enhanced as $\xi$ increases, particularly for $\xi \ge 0.02$. According to the expression for $c_S$ in Eq. \eqref{ssr}, a larger value of $\xi$ results in more pronounced oscillations in $\phi_k$. In the \textbf{RD} and \textbf{MD} eras, the case with $\xi = 0.02$ markedly amplifies the amplitude of $\phi_k$, serving as a distinguishing feature between the standard and non-trivial scenarios. For other values of $\xi$, similar oscillatory behavior is observed, along with an overall increase in the amplitude of $\phi_k$. The evolutionary trends remain consistent across different values of $\xi$.

In summary, a non-trivial sound speed increases the amplitude of $r_k$ and $\phi_k$ during inflation, radiation-dominated, and matter-dominated epochs. It may also induce oscillations in $\phi_k$ and $r_k$, particularly during inflation.


\subsection{Evolution of Krylov complexity and Krylov entropy}
\label{evolution of Krylov complexity and Krylov entropy}
\begin{figure}
	\centering
	\includegraphics[width=.4\textwidth]{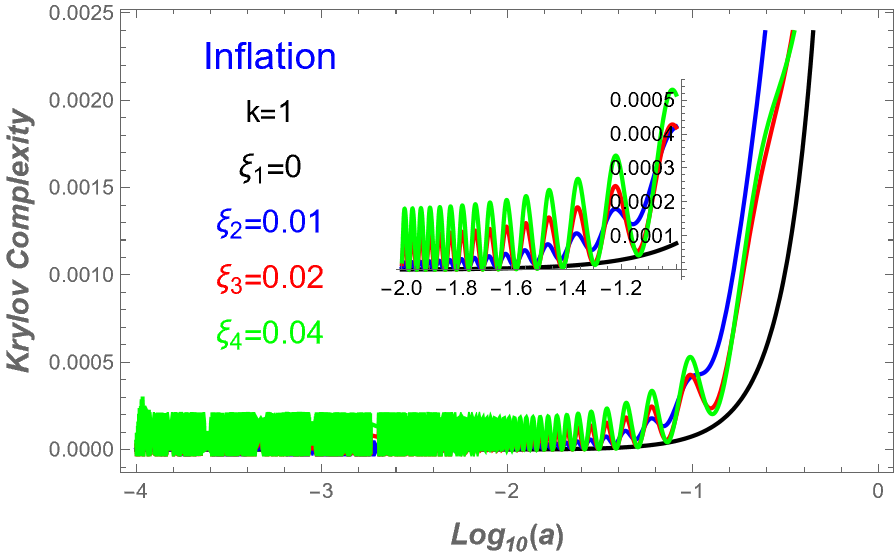}\\
	\qquad
	\includegraphics[width=.4\textwidth]{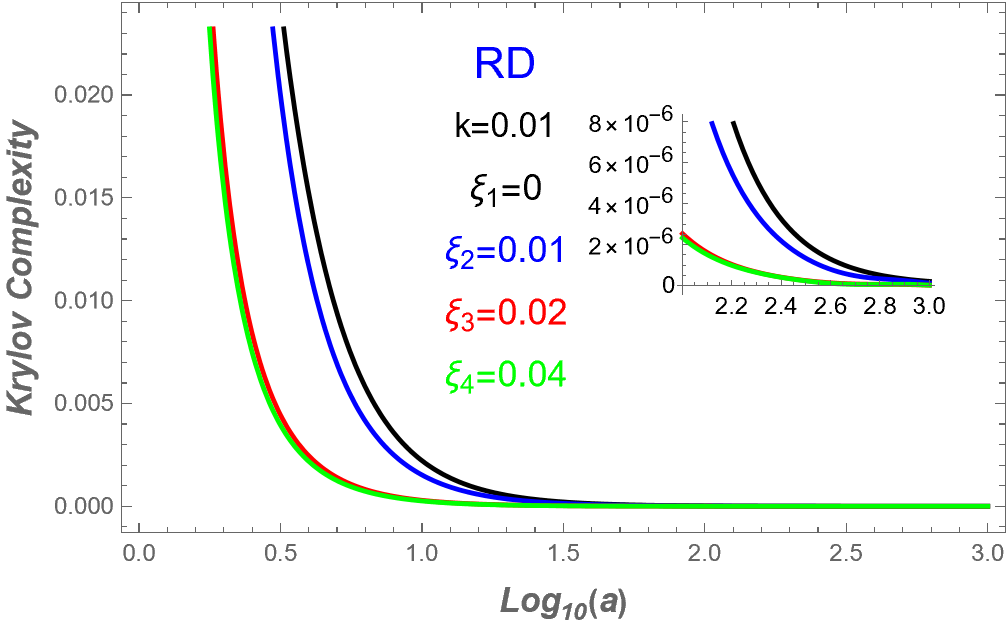}
	\qquad
	\includegraphics[width=.4\textwidth]{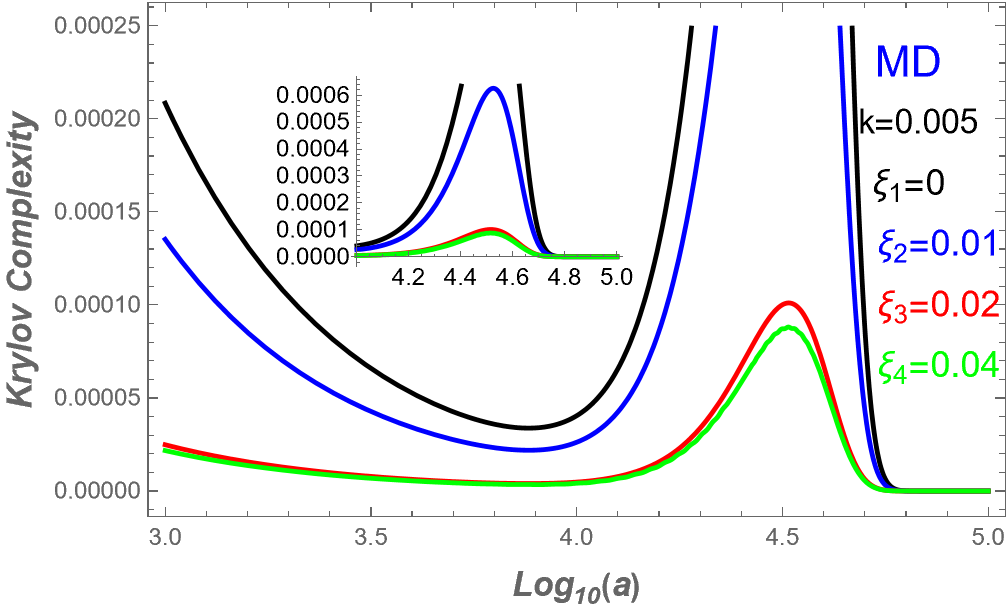}
	\caption{The numerical of Krylov complexity in terms of $\log_{10}a$ for three different periods (inflation, RD, and MD), where we set $H_{0}=1$, $k=1$ in inflation, $k=0.01$ at RD and $k=0.005$ at MD for simplicity.}
	\label{fig: op s KC}
\end{figure}

\begin{figure}
	\centering
	\includegraphics[width=.4\textwidth]{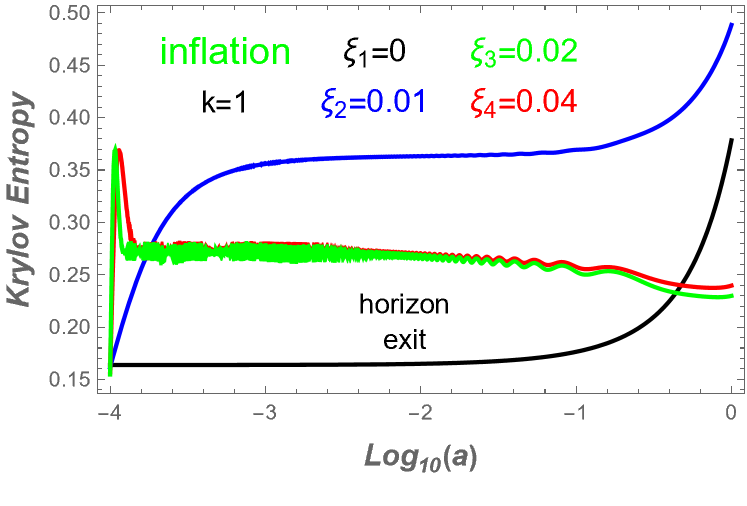}\\
	\qquad
	\includegraphics[width=.4\textwidth]{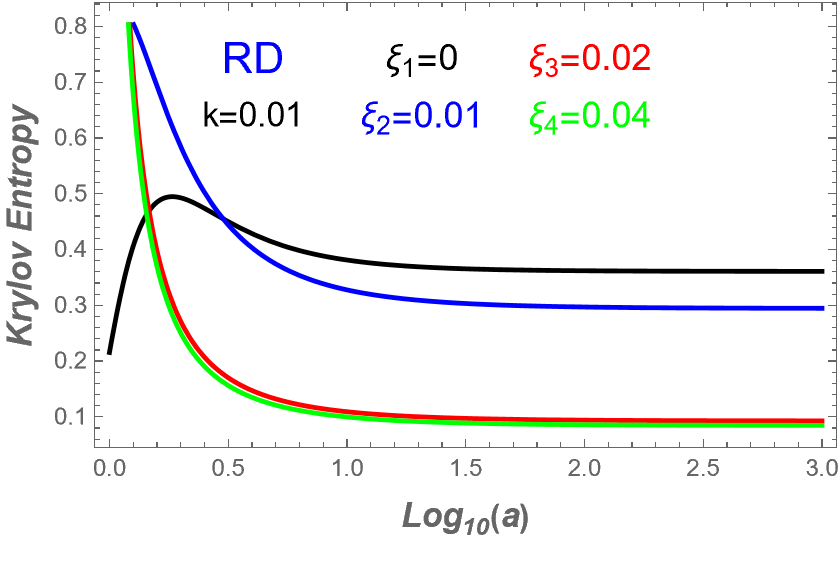}
	\qquad
	\includegraphics[width=.4\textwidth]{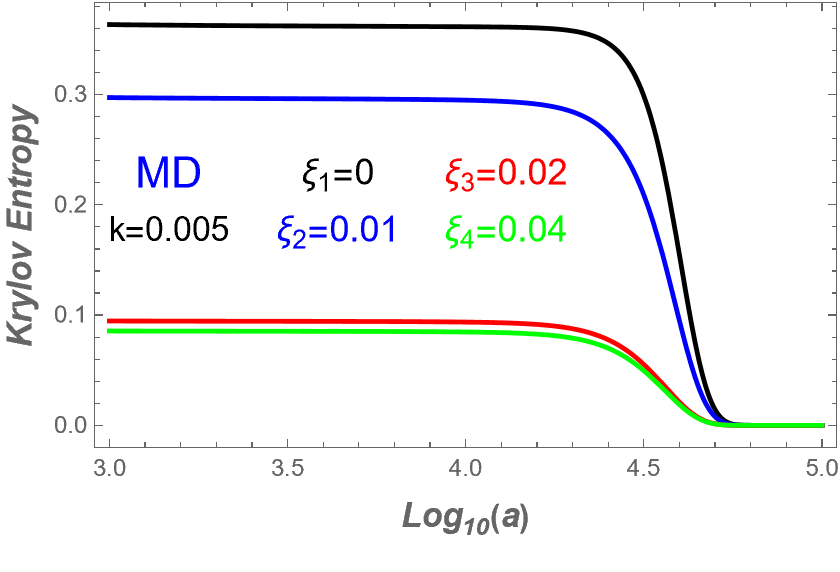}
	\caption{The numerical of Krylov entropy in terms of $\log_{10}a$ for three different periods (inflation, RD, and MD), where we set $H_{0}=1$, $k=1$ in inflation, $k=0.01$ at RD and $k=0.005$ at MD for simplicity.}
	\label{fig: op s KE}
\end{figure}

In this section, we use the evolution of $r_k$ to numerically simulate Krylov complexity and Krylov entropy. The definitions of Krylov complexity and Krylov entropy are provided below, 
\begin{equation}
\begin{split}
\begin{cases}
&K=\sum_{n=0}^{\infty}n|\phi_n|^2\\&S_k=-\sum_{n=0}^{\infty}|\phi|^2\ln{|\phi|^2}\label{5.1}
\end{cases}
\end{split}
\end{equation}
Based on \eqref{wave function1}, we could derive the formula of Krylov complexity and Krylov entropy, 
\begin{equation}
\begin{split}
K=\frac{(|1-u_1^{2}|\tanh^2 r_k)\mbox{sech}^2(r_k)}{[1+2u_2\tanh{r_k}+(u_2^2-|1-u_1^2|)\tanh{r_k}]^2}  
\end{split}
\label{krylov complexity}  
\end{equation}
\begin{equation}
\begin{split}
S_k&=\frac{\rm{sech^2r_k}}{A^2}\bigg[\ln{(1+u_2\tanh{r_k})^2}(1+u_2\tanh{r_k})^2-\big((1+u_2\tanh{r_k})^2\\&-|1-u_1^2|\big)\tanh^2{r_k})\ln{\rm{sech^2r_k}}-|1-u_1^2|\tanh^2{r_k}\ln{|1-u_1^2|}-|1-u_1^2|\tanh^2{r_k}\ln{\tanh^2{r_k}}\bigg]  
\end{split}
\label{krylov entropy}
\end{equation}
where $A=1+2u_2\tanh r_k+(u_2^2-|1-u_1^2|)\tanh^2 r_k$. In light of \eqref{krylov complexity} and \eqref{krylov entropy}, we could obtain the key results of this paper, as shown in Figs. \ref{fig: op s KC} and \ref{fig: op s KE}.

Fig.~\ref{fig: op s KC} clearly demonstrates distinct evolutionary patterns for Krylov complexity across different cosmological epochs:

\textbf{Inflationary Era:} The Krylov complexity exhibits characteristic exponential growth, consistent with the findings of \cite{Zhai:2024odw}. While larger values of $\xi$ introduce more pronounced oscillatory behavior, the overall exponential growth trend remains consistent across different $\xi$ values. All cases show growth amplitudes of comparable order of magnitude.

\textbf{RD Era:} The Krylov complexity decays to negligible values (approximately $\mathcal{O}(10^{-6})$). Notably, larger $\xi$ values correlate with faster decay rates compared to smaller $\xi$ cases.

\textbf{MD Era:} The evolution displays more complex behavior, featuring a distinct peak near $y = 4.5$. The standard case ($\xi = 0$) produces the most prominent peak, while increasing $\xi$ values result in progressively smaller peaks. Interestingly, although $\xi = 0.02$ produces significant differences in other observables (particularly regarding $\phi_k$), its impact on Krylov complexity evolution appears less pronounced.

The evolution of Krylov entropy across different cosmological epochs is presented in Fig.~\ref{fig: op s KE}, revealing distinct behavior compared to Krylov complexity as follows: 

\textbf{Inflationary Era:}
The standard single-field inflation scenario ($\xi = 0$) exhibits sustained exponential growth in Krylov entropy. For $\xi = 0.01$, the entropy increases until reaching a quasi-constant value, then resumes growth after $\log_{10} a > -1$. A notable transition occurs at $\xi = 0.02$, where the entropy develops a distinct peak prior to horizon exit before saturating to constant values―a behavior also observed for $\xi = 0.04$. This sharp transition establishes $\xi = 0.02$ as a critical value for distinguishing between standard and non-trivial sound speed inflationary models.

\textbf{RD Era:}
The standard case shows a monotonic increase toward a constant saturation value. In contrast, all non-zero $\xi$ cases display similar decaying trends, eventually stabilizing at constant values.

\textbf{MD Era:}
Both standard and non-trivial cases initially maintain different constant values, followed by a decay to negligible magnitudes ($\mathcal{O}(10^{-6})$).


In this section, we employ numerical methods to investigate the evolution of Krylov complexity and Krylov entropy with respect to the parameters $\phi_k$ and $r_k$. Recall that Krylov complexity quantifies the growth of an operator and typically exhibits exponential growth before saturating at a maximum value. This behavior indicates that the dynamical system is maximally chaotic, with Lanczos coefficients proportional to the principal quantum number $n$. Based on this property, we can conclude that the inflationary universe constitutes a maximally chaotic system regardless of whether $\xi$ vanishes or not. Physically, this implies that both standard single-field inflation and models with non-trivial sound speed are maximally chaotic systems, as demonstrated in \cite{Li:2024kfm,Li:2024iji,Li:2024ljz}. However, the situation becomes more complex during the RD and MD epochs. Although the Lanczos coefficients remain proportional to $n$ as given in Eq.~\eqref{eq bn1}, a consequence of the Hamiltonian structure, the overall behavior differs. The detailed trends are illustrated in Fig.~\ref{fig: op s KC}.

Krylov entropy quantifies the disorder or spread of an operator's evolution within the Krylov space \cite{He:2024xjp}, measuring the "disorder" of the Lanczos coefficients. This paper demonstrates its utility in distinguishing between standard single-field inflation and models with non-trivial sound speed. Our analysis reveals that Krylov entropy acts as a powerful diagnostic tool to distinguish between standard single-field inflation and models with a non-trivial sound speed ($c_S \neq 1$). A key finding is the significant impact of the parameter $\phi_k$, which sets the scale of the feature in the sound speed---on the Krylov entropy. Numerical simulations reveal that $\xi = 0.02$ acts as a critical threshold for the behavior of Krylov entropy. At this value, the oscillatory features introduced by the Sound Speed Resonance (SSR) become dominant enough to produce a distinct peak in entropy before horizon exit. While the specific numerical value may shift slightly with different choices of momentum $k$ (we have set $H_0=1$ for the entire early universe), the emergence of this 'peak-and-saturate' behavior is a robust feature of the SSR model that distinguishes it from the monotonic growth seen in standard single-field inflation in inflationary period. This distinction is not limited to inflation; it is also evident in the subsequent RD and MD eras. The sensitivity of Krylov entropy to the feature scale $\phi_k$ establishes it as a robust criterion for discriminating between standard single-field inflation and models with non-trivial kinetic terms characterized by $c_S$.

\section{Summary and discussion}
\label{summary and discussion}
In this work, we have systematically investigated the Krylov complexity for the whole early universe, including inflation, RD and MD, which is under the framework of non-trivial sound speed. Our analysis could recover most quantum gravitational frameworks, for instance, k-inflation, DBI inflation via string theory and the effective field theory could all produce the non-trivial sound speed. 

Given the inherent open-system nature of the universe, we employ an open-system approach within the generalized Lanczos algorithm, where the generalized Krylov basis is constructed via the Arnoldi iteration as detailed in Sec.\ref{lanczos algorithm}. In this framework, the Lindbladian takes an upper Heisenberg form through the Arnoldi process, in which the diagonal elements $h_{n,n}$ correspond to the dissipative part $c_n$ (open-system contribution), while the off-diagonal elements $h_{n,n-1}$ correspond to the $b_n$ term (closed-system contribution).

Using the specific Hamiltonian given in \eqref{final hamiltonian}, we can characterize the statistical properties of the early universe via the Lanczos coefficient \eqref{eq bn1} and the dissipative coefficient \eqref{eq mu1 mu2}. From \eqref{eq bn1}, it follows that the early universe constitutes a maximally chaotic system, as indicated by the linear growth $b_n \propto n$. Accordingly, $b_n$ can be related to the Lyapunov exponent through $\lambda = b_2$, following the argument in \cite{Parker:2018yvk}. This connection allows us to analyze the statistical behavior of the universe systematically. 

Fig. \ref{fig: op s bn} shows that the overall trend of the Lanczos coefficient remains similar across different cosmological epochs, although non-trivial sound speed introduces more pronounced oscillations during inflation. Meanwhile, Fig. \ref{fig: op s u2} reveals that inflation behaves as a strongly dissipative regime, whereas the radiation-dominated (RD) and matter-dominated (MD) eras exhibit weak dissipation. Importantly, the presence of a non-trivial sound speed does not alter the fundamental dissipative character of the early universe.

To determine the evolution of Krylov complexity, the key step involves numerically solving for two parameters: $\phi_k$ and $r_k$. In this work, we present the evolution equations for these parameters under non-trivial sound speed for the firs time, derived from the Schrödinger equation \eqref{schrodinger equation} as given in Eqs. \eqref{rk evolution} and \eqref{phik evolution}. Using these equations, Fig. \ref{fig: op s rk} clearly illustrates the behavior of $r_k$, showing that greater deviations from the standard case (denoted by $\xi$) lead to larger amplitudes of $r_k$ during its evolution. As for $\phi_k$, more pronounced distinctions arise between the standard case and non-trivial sound speed models, as seen in Figs. \ref{fig: op s phi} and \ref{fig: op s phi1}, where $\xi = 0.02$ serves as a critical value distinguishing the two scenarios. 

Fig. \ref{fig: op s KC} clearly shows the evolution of Krylov complexity: it grows exponentially during inflation, decays toward a constant in the radiation-dominated (RD) era, and exhibits a peak in the matter-dominated (MD) epoch. While all cases follow a similar trend, a key difference from \cite{Parker:2018yvk} is that, although the early universe remains a chaotic system, the Krylov complexity does not saturate to a fixed value, both for non-trivial sound speed and standard cases, consistent with our previous findings \cite{Li:2024kfm}. Unlike many finite-dimensional chaotic systems where Krylov complexity eventually reaches a constant saturation value, our results show different trend. This distinction arises from the infinite-dimensional nature of the Hilbert space associated with the continuous mode evolution in an expanding universe.

For completeness, we also analyze the Krylov entropy, which quantifies the disorder of operator evolution in the Arnoldi space (a generalized Krylov space) and serves as a complement to Krylov complexity. Fig. \ref{fig: op s KE} reveals distinct evolutionary behavior of the Krylov entropy under non-trivial sound speed, particularly for $\xi = 0.02$ and larger values. During inflation, the standard case exhibits exponential growth, consistent with \cite{Zhai:2024odw}, while for $\xi = 0.02$, the entropy peaks before horizon exit and then saturates to a constant. In the RD era, the overall trend for $\xi = 0.02$ is also altered, converging again to a constant. During MD, the behavior is similar across cases, with the value of $\xi$ mainly affecting the amplitude of the Krylov entropy.

In summary, our analysis reveals that the early universe constitutes a maximally chaotic system, as indicated by the scaling $b_n \propto n$. However, unlike typical chaotic systems, the Krylov complexity does not saturate to a finite value. This behavior can be attributed to the expansion of spacetime in the standard inflationary scenario. Our work not only confirms this point but also offers a new quantum-informatic perspective for discriminating between standard single-field inflation and models with non-trivial sound speed. Furthermore, by employing the open quantum system approach to construct the basis for curvature perturbations, it is natural to explore decoherence effects within this framework \cite{Bhattacharjee:2022lzy}. We have previously examined the power spectrum using an open two-mode squeezed state formalism \cite{Zhai:2025abc}, and future work may incorporate non-Gaussianities. While the present analysis applies broadly to single-field inflation with non-trivial sound speed, it can be extended to multi-field inflationary scenarios \cite{Liu:2019xhn,Liu:2020zzv,Zhang:2022bde,Liu:2021rgq,Liu:2020zlr} and modified gravity theories, such as $f(R)$ gravity \cite{Liu:2018hno,Liu:2018htf}.

 \section*{Acknowledgements}
LHL and SHL are funded by National Natural Science Foundation of China (NSFC) with grant NO. 12165009, Hunan Natural Science Foundation with grant NO. 2023JJ30487 and NO. 2022JJ40340. HQZ and PZH is funded by NSFC with grant NO. 12175008. 

\section*{Data availability Statement}
Data sharing not applicable to this article as no datasets were generated or analysed during the current study.

\appendix
\section{The calculation of $r_k$ and $\phi_k$ within the method of open system}\label{Appendix A}
In this appendix, we will give the detailed calculation for the evolution equation of $r_k$ and $\phi_k$. First, we could write the Hamiltonian in the following form,
\begin{equation}
\begin{split}
\hat{H_k}&=[(\frac{a^{2}V_{\phi\phi}}{2k}+\frac{z'}{z}i-\frac{k}{2}+\frac{kc_S^2}{2}){\hat{c}^{\dagger} _{\vec{k}}}{\hat{c}^{\dagger} _{\vec{-k}}}+(\frac{a^2V_{\phi\phi}}{2k}-\frac{z'}{z}i-\frac{k}{2}+\frac{kc_S^2}{2}){\hat{c} _{\vec{k}}}{\hat{c}_{\vec{-k}}}\\&+(\frac{a^2V_{\phi\phi}}{2k}+\frac{k}{2}+\frac{kc_S^2}{2})({\hat{c} _{\vec{k}}}{\hat{c}^{\dagger} _{\vec{k}}}+{\hat{c} _{\vec{-k}}}{\hat{c}^{\dagger} _{\vec{-k}}})]
\end{split}
\end{equation}
To facilitate computation, the Hamiltonian operator is decomposed into three distinct components, 
\[\hat{H}=\hat{H}_{1}+\hat{H}_{2}+\hat{H}_{3}\] 
where 
\[H_{1}=(\frac{a^2V_{\phi\phi}}{2k}+\frac{z'}{z}i-\frac{k}{2}+\frac{kc_S^2}{2}){\hat{c}^{\dagger} _{\vec{k}}}{\hat{c}^{\dagger} _{\vec{-k}}}\]
\[H_{2}=(\frac{a^2V_{\phi\phi}}{2k}-\frac{z'}{z}i-\frac{k}{2}+\frac{kc_S^2}{2}){\hat{c} _{\vec{k}}}{\hat{c}_{\vec{-k}}}\]
\[H_{3}=(\frac{a^2V_{\phi\phi}}{2k}+\frac{k}{2}+\frac{kc_S^2}{2})({\hat{c} _{\vec{k}}}{\hat{c}^{\dagger} _{\vec{k}}}+{\hat{c}^{\dagger}_{-\vec{k}}}{\hat{c} _{-\vec{k}}})\]
Then, making use of \eqref{wave function1} and Schr\"{o}dinger equation \eqref{schrodinger equation}, we could caclulate $H_1$, $H_2$ and $H_3$, respectively, 
\begin{equation}
\begin{split}
&\hat{H}_{1}| \varphi\rangle=(\frac{a^2V_{\phi\phi}}{2k}+\frac{z'}{z}i-\frac{k}{2}+\frac{kc_S^2}{2})(\frac{\rm{sech}(r_k)}{1+u_2\tanh(r_k)})\\&\sum_{n=0}^\infty (-1)^n |1-u_1^2|^\frac{n}{2}e^{2in\phi_k}(\frac{\tanh(r_k)}{1+u_2\tanh(r_k)})^n {\hat{c}^{\dagger} _{\vec{k}}}{\hat{c}^{\dagger} _{\vec{-k}}}|{n,n}\rangle_{\vec{k},-\vec{k}}
\\&=(\frac{a^2V_{\phi\phi}}{2k}+\frac{z'}{z}i-\frac{k}{2}+\frac{kc_S^2}{2})(\frac{\rm{sech}(r_k)}{1+u_2\tanh(r_k)})\\&\sum_{n=0}^\infty (-1)^n |1-u_1^2|^\frac{n}{2}e^{2in\phi_k}(\frac{\tanh(r_k)}{1+u_2\tanh(r_k)})^n (n+1)| {n+1,n+1}\rangle_{\vec{k},-\vec{k}}
\\&=(\frac{a^2V_{\phi\phi}}{2k}+\frac{z'}{z}i-\frac{k}{2}+\frac{kc_S^2}{2})(\frac{\rm{sech}(r_k)}{1+u_2\tanh(r_k)})\\&\sum_{n=1}^\infty (-1)^{n-1} |1-u_1^2|^\frac{n-1}{2}e^{2i(n-1)\phi_k}(\frac{\tanh(r_k)}{1+u_2\tanh(r_k)})^{(n -1)}n| {n,n}\rangle_{\vec{k},-\vec{k}}
\end{split}
\label{H1}
\end{equation}

\begin{equation}
\begin{split}
&\hat{H}_{2}|\varphi\rangle=(\frac{a^2V_{\phi\phi}}{2k}-\frac{z'}{z}i-\frac{k}{2}+\frac{kc_S^2}{2})(\frac{\rm{sech}(r_k)}{1+u_2\tanh(r_k)})\\&\sum_{n=0}^\infty (-1)^n |1-u_1^2|^\frac{n}{2}e^{2in\phi_k}(\frac{\tanh(r_k)}{1+u_2\tanh(r_k)})^n {\hat{c} _{\vec{k}}}{\hat{c}_{\vec{-k}}}| {n,n}\rangle_{\vec{k},-\vec{k}}
\\&=(\frac{a^2V_{\phi\phi}}{2k}-\frac{z'}{z}i-\frac{k}{2}+\frac{kc_S^2}{2})(\frac{\rm{sech}(r_k)}{1+u_2\tanh(r_k)})\\&\sum_{n=0}^\infty (-1)^n |1-u_1^2|^\frac{n}{2}e^{2in\phi_k}(\frac{\tanh(r_k)}{1+u_2\tanh(r_k)})^n n| {n-1,n-1}\rangle_{\vec{k},-\vec{k}}
\\&=(\frac{a^2V_{\phi\phi}}{2k}-\frac{z'}{z}i-\frac{k}{2}+\frac{kc_S^2}{2})(\frac{\rm{sech}(r_k)}{1+u_2\tanh(r_k)})[(-1) |1-u_1^2|^\frac{1}{2}e^{2i\phi_k}(\frac{\tanh(r_k)}{1+u_2\tanh(r_k)}) | {0,0}\rangle_{\vec{k},-\vec{k}}\\&+\sum_{n=1}^\infty (-1)^{(n+1)}|1-u_1^2|^\frac{n+1}{2}e^{2i(n+1)\phi_k}(\frac{\tanh(r_k)}{1+u_2\tanh(r_k)})^{(n+1)}(n+1)| {n,n}\rangle_{\vec{k},-\vec{k}}]
\end{split}
\label{H2}
\end{equation}

\begin{equation}
\begin{split}
&\hat{H}_{3}|\varphi\rangle=(\frac{a^2V_{\phi\phi}}{2k}+\frac{k}{2}+\frac{kc_S^2}{2})(\frac{\rm{sech}(r_k)}{1+u_2\tanh(r_k)})\\&\sum_{n=0}^\infty (-1)^n |1-u_1^2|^\frac{n}{2}e^{2in\phi_k}(\frac{\tanh(r_k)}{1+u_2\tanh(r_k)})^n ({\hat{c} _{\vec{k}}}{\hat{c}^{\dagger} _{\vec{k}}}+{\hat{c} _{-\vec{k}}}{\hat{c}^{\dagger} _{-\vec{k}}})| {n,n}\rangle_{\vec{k},-\vec{k}}
\\&=(\frac{a^2V_{\phi\phi}}{2k}+\frac{k}{2}+\frac{kc_S^2}{2})(\frac{\rm{sech}(r_k)}{1+u_2\tanh(r_k)})\\&\sum_{n=0}^\infty (-1)^n |1-u_1^2|^\frac{n}{2}e^{2in\phi_k}(\frac{\tanh(r_k)}{1+u_2\tanh(r_k)})^n (2n+1)| {n,n}\rangle_{\vec{k},-\vec{k}}
\\&=(\frac{a^2V_{\phi\phi}}{2k}+\frac{k}{2}+\frac{kc_S^2}{2})(\frac{\rm{sech}(r_k)}{1+u_2\tanh(r_k)})\\&[| {0,0}\rangle_{\vec{k},-\vec{k}}+\sum_{n=0}^\infty (-1)^n |1-u_1^2|^\frac{n}{2}e^{2in\phi_k}(\frac{\tanh(r_k)}{1+u_2\tanh(r_k)})^n (2n+1)| {n,n}\rangle_{\vec{k},-\vec{k}}]
\end{split}
\label{H3}
\end{equation}
For the following calculations, we need to compare the excited state and the ground state for the both sides of  Schr\"{o}dinger equation \eqref{schrodinger equation}, that is reason why we list the ground state and excited state for \eqref{H1}, \eqref{H2} and \eqref{H3}. We outline the ground state and excited states for the Hamiltonian operator's side. First, it is the ground state, 
\begin{equation}
\begin{split}
&(\frac{\rm{sech}(r_k)}{1+u_2\tanh(r_k)})[(\frac{a^2V_{\phi\phi}}{2k}+\frac{k}{2}+\frac{kc_S^2}{2})-(\frac{a^2V_{\phi\phi}}{2k}-\frac{z'}{z}i-\frac{k}{2}+\frac{kc_S^2}{2})\\&\times|1-u_1^2|^{\frac{1}{2}}e^{2i\phi_k}(\frac{\tanh(r_k)}{1+u_2\tanh(r_k)})] | {0,0}\rangle_{\vec{k},-\vec{k}}
\end{split}
\end{equation}
Second, it is the excited state without $n$-dependent terms;
\begin{equation}
\begin{split}
&(\frac{\rm{sech}(r_k)}{1+u_2\tanh(r_k)})[(\frac{a^2V_{\phi\phi}}{2k}+\frac{k}{2}+\frac{kc_S^2}{2})-(\frac{a^2V_{\phi\phi}}{2k}-\frac{z'}{z}i-\frac{k}{2}+\frac{kc_S^2}{2})\\\times&|1-u_1^2|^{\frac{1}{2}}e^{2i\phi_k}(\frac{\tanh(r_k)}{1+u_2\tanh(r_k)})] \sum_{n=0}^\infty (-1)^n |1-u_1^2|^\frac{n}{2}e^{2in\phi_k}\\&\times(\frac{\tanh(r_k)}{1+u_2\tanh(r_k)})^n| {n,n}\rangle_{\vec{k},-\vec{k}}
\end{split}
\end{equation}
Finally, it is the state with $n$-dependent terms as following, 
\begin{equation}
\begin{split}
&=(\frac{\rm{sech}(r_k)}{1+u_2\tanh(r_k)})[(\frac{a^2V_{\phi\phi}}{k}+k+kc_S^2)-(\frac{a^2V_{\phi\phi}}{2k}-\frac{z'}{z}i-\frac{k}{2}+\frac{kc_S^2}{2})\\&|1-u_1^2|^{\frac{1}{2}}e^{2i\phi_k}(\frac{\tanh(r_k)}{1+u_2\tanh(r_k)})-(\frac{a^2V_{\phi\phi}}{2k}+\frac{z'}{z}i-\frac{k}{2}+\frac{kc_S^2}{2})|1-u_1^2|^{-\frac{1}{2}}e^{-2i\phi_k}\\&(\frac{\tanh(r_k)}{1+u_2\tanh(r_k)})^{-1}]\sum_{n=0}^\infty (-1)^n |1-u_1^2|^\frac{n}{2}e^{2in\phi_k}\\&(\frac{\tanh(r_k)}{1+u_2\tanh(r_k)})^nn| {n,n}\rangle_{\vec{k},-\vec{k}}
\end{split}
\end{equation}
Then, we will explicitly derivative with the conformal time for the OTMSS whose calculation process is as follows, 
\begin{equation}
\begin{split}
& i\frac{d}{d\eta}|{\varphi}\rangle=i\frac{d}{d\eta}[\frac{\rm{sechr_k}}{1+u_2\tanh{r_k}}\sum^{\infty}_{n=0}(-1)^n|1-u_1^2|^{\frac{n}{2}}e^{2in\phi_k}(\frac{\tanh(r_k)}{1+u_2\tanh(r_k)})^n| {n,n}\rangle_{\vec{k},-\vec{k}}]\\&=i[\frac{-\rm{sechr_k\tanh{r_k(1+u_2\tanh{r_k})r'_k-\rm{sechr_k(u'_2\tanh{r_k}+u_2r'_k\rm{sech^2r_k})}}}}{(1+u_2\tanh{r_k})^2}\\&\sum^{\infty}_{n=0}(-1)^n|1-u_1^2|^{\frac{n}{2}}e^{2in\phi_k}(\frac{\tanh(r_k)}{1+u_2\tanh(r_k)})^n| {n,n}\rangle_{\vec{k},-\vec{k}}\\&+\frac{\rm{sechr_k}}{1+u_2\tanh{r_k}}\frac{d}{d\eta}\sum^{\infty}_{n=0}(-1)^n|1-u_1^2|^{\frac{n}{2}}e^{2in\phi_k}(\frac{\tanh(r_k)}{1+u_2\tanh(r_k)})^n| {n,n}\rangle_{\vec{k},-\vec{k}}]
\\&=i[\frac{(-\sinh{2r_k-2u_2\cosh^2{r_k}})r'_k-u'_2\sinh{2r_k}}{2\cosh^3{r_k}(1+u_2\tanh{r_k})^2}\sum^{\infty}_{n=0}(-1)^n|1-u_1^2|^{\frac{n}{2}}e^{2in\phi_k}(\frac{\tanh(r_k)}{1+u_2\tanh(r_k)})^n| {n,n}\rangle_{\vec{k},-\vec{k}}]
\\&+i\frac{\rm{sechr_k}}{1+u_2\tanh{r_k}}\sum^{\infty}_{n=0}(-1)^n[\frac{n}{2}|1-u_1^2|^{\frac{n}{2}-1}|1-u_1^2|'e^{2in\phi_k}(\frac{\tanh(r_k)}{1+u_2\tanh(r_k)})^n\\&+|1-u_1^2|^{\frac{n}{2}}2in\phi'_ke^{2in\phi_k}(\frac{\tanh(r_k)}{1+u_2\tanh(r_k)})^n\\&+|1-u_1^2|^{\frac{n}{2}}e^{2in\phi_k}n(\frac{\tanh(r_k)}{1+u_2\tanh(r_k)})^{(n-1)}\frac{r'_k\rm{sechr^2_k-u'_2\tanh^2{r_k}}}{(1+u_2\tanh{r_k})^2}]| {n,n}\rangle_{\vec{k},-\vec{k}}
\\&=i[\frac{(-\sinh{2r_k-2u_2\cosh^2{r_k}})r'_k-u'_2\sinh{2r_k}}{2\cosh^3{r_k}(1+u_2\tanh{r_k})^2}\sum^{\infty}_{n=0}(-1)^n|1-u_1^2|^{\frac{n}{2}}e^{2in\phi_k}(\frac{\tanh(r_k)}{1+u_2\tanh(r_k)})^n| {n,n}\rangle_{\vec{k},-\vec{k}}]
\\&+i\frac{\rm{sechr_k}}{1+u_2\tanh{r_k}}\sum^{\infty}_{n=0}(-1)^n|1-u_1^2|^{\frac{n}{2}}e^{2in\phi_k}(\frac{\tanh(r_k)}{1+u_2\tanh(r_k)})^n\\&[2i\phi'_k+\frac{2r'_k}{\sin{2r_k(1+u_2\tanh{r_k})}}+\frac{|1-u_1^2|'}{2|1-u_1^2|}-\frac{u'_2\tanh{r_k}}{1+u_2\tanh{r_k}}]| {n,n}\rangle_{\vec{k},-\vec{k}}
\end{split}
\end{equation}
Following the similar procedure, it can also be categorized into the ground state, excited states without $n$-dependent term, and excited states with $n$-dependent term. First, we list the ground state as follows, 

\begin{equation}
\begin{split}
i\frac{(\sinh{2r_k}+2u_2\cosh^2{r_k})r'_k+u'_2\sin{2r_k}}{2\cosh^3{r_k}(1+u_2\tanh{r_k})^2}|{0,0}\rangle_{\vec{k},-\vec{k}}
\end{split}
\end{equation}
Then, the excited state without $n$-dependent term is as follows, 
\begin{equation}
\begin{split}
-i\frac{(\sinh{2r_k}+2u_2\cosh^2{r_k})r'_k+u'_2\sin{2r_k}}{2\cosh^3{r_k}(1+u_2\tanh{r_k})^2}\sum_{n=0}^\infty  |1-u_1^2|^\frac{n}{2}(-\frac{\tanh(r_k)e^{2i\phi_k}}{1+u_2\tanh(r_k)})^n| {n,n}\rangle_{\vec{k},-\vec{k}}
\end{split}
\end{equation}
Thirdly, it is also the excited term with $n$-dependent term 

\begin{equation}
\begin{split}
&(\frac{\rm{sech}(r_k)}{1+u_2\tanh(r_k)})[-2\phi'_k+i\frac{2r'_k}{\sin{2r_k(1+u_2\tanh{r_k})}}+i\frac{|1-u^2_1|'}{2|1-u^2_1|}\\&-i\frac{u'_2\tanh{r_k}}{{(1+u_2\tanh{r_k})}}]\sum_{n=0}^\infty (-1)^n |1-u_1^2|^\frac{n}{2}e^{2in\phi_k}(\frac{\tanh(r_k)}{1+u_2\tanh(r_k)})^nn| {n,n}\rangle_{\vec{k},-\vec{k}}
\end{split}
\end{equation}
Through the above calculations, we could observe that the ground state and the excited state without $n$-dependent term will produce the evolution of $r_k$. As for the evolution equation of $\phi_k$, it is hidden in the real part of excited state with $n$-dependent term. After some involved calculations, we could summarize the final formula of $r_k$ and $\phi_k$ as follows,
\begin{equation}
\begin{split}
r'_k&=\frac{|1-u_1^2|^{\frac{1}{2}}\sinh{2r_k}[(\frac{a^2V_{\phi\phi}}{2k}-\frac{k}{2}+\frac{kc_S^2}{2})\sin{2\phi_k}-\frac{z'}{z}\cos{2\phi_k}]-u'_2\sinh{2r_k}}{\sin{2r_k}+2u_2\cosh^2{r_k}}\\
\phi'_k=&-\frac{1}{2}(k+kc^2_S+\frac{a^2V_{\phi\phi}}{k})+\frac{1}{2}[(\frac{a^2V_{\phi\phi}}{2k}-\frac{k}{2}+\frac{kc^2_S}{2})\cos{2\phi_k}+\frac{z'}{z}\sin{2\phi_k}]\\&(|1-u_1^2|^{-\frac{1}{2}}(u_2+\coth{r_k})+|1-u_1^2|^{\frac{1}{2}}\frac{\tanh{r_k}}{1+u_2\tanh{r_k}}).
\end{split}
\label{evolution of r and phi}
\end{equation}


\section*{References}
\begin{thebibliography}{99}
	

\bibitem{Agullo:2022ttg}
I.~Agullo, B.~Bonga and P.~R.~Metidieri,
JCAP \textbf{09} (2022), 032
doi:10.1088/1475-7516/2022/09/032
[arXiv:2203.07066 [gr-qc]].


\bibitem{Unruh:1976db}
W.~G.~Unruh,
Phys. Rev. D \textbf{14} (1976), 870
doi:10.1103/PhysRevD.14.870




\bibitem{Grishchuk:1990bj}
L.~P.~Grishchuk and Y.~V.~Sidorov,
Phys. Rev. D \textbf{42} (1990), 3413-3421
doi:10.1103/PhysRevD.42.3413


\bibitem{Parker:2018yvk}
D.~E.~Parker, X.~Cao, A.~Avdoshkin, T.~Scaffidi and E.~Altman,
Phys. Rev. X \textbf{9} (2019) no.4, 041017
doi:10.1103/PhysRevX.9.041017
[arXiv:1812.08657 [cond-mat.stat-mech]].


\bibitem{Rabinovici:2020ryf}
E.~Rabinovici, A.~S{\'a}nchez-Garrido, R.~Shir and J.~Sonner,
JHEP \textbf{06} (2021), 062
doi:10.1007/JHEP06(2021)062
[arXiv:2009.01862 [hep-th]].


\bibitem{Jian:2020qpp}
S.~K.~Jian, B.~Swingle and Z.~Y.~Xian,
JHEP \textbf{03} (2021), 014
doi:10.1007/JHEP03(2021)014
[arXiv:2008.12274 [hep-th]].

\bibitem{He:2022ryk}
S.~He, P.~H.~C.~Lau, Z.~Y.~Xian and L.~Zhao,
JHEP \textbf{12} (2022), 070
doi:10.1007/JHEP12(2022)070
[arXiv:2209.14936 [hep-th]].


\bibitem{Cao:2020zls}
X.~Cao,
J. Phys. A \textbf{54} (2021) no.14, 144001
doi:10.1088/1751-8121/abe77c
[arXiv:2012.06544 [cond-mat.stat-mech]].



\bibitem{Trigueros:2021rwj}
F.~B.~Trigueros and C.~J.~Lin,
SciPost Phys. \textbf{13} (2022) no.2, 037
doi:10.21468/SciPostPhys.13.2.037
[arXiv:2112.04722 [cond-mat.dis-nn]].


\bibitem{Heveling:2022hth}
R.~Heveling, J.~Wang and J.~Gemmer,
Phys. Rev. E \textbf{106} (2022) no.1, 014152
doi:10.1103/PhysRevE.106.014152
[arXiv:2203.00533 [cond-mat.stat-mech]].

\bibitem{Dymarsky:2021bjq}
A.~Dymarsky and M.~Smolkin,
Phys. Rev. D \textbf{104} (2021) no.8, L081702
doi:10.1103/PhysRevD.104.L081702
[arXiv:2104.09514 [hep-th]].


\bibitem{Caputa:2021ori}
P.~Caputa and S.~Datta,
JHEP \textbf{12} (2021), 188
[erratum: JHEP \textbf{09} (2022), 113]
doi:10.1007/JHEP12(2021)188
[arXiv:2110.10519 [hep-th]].


\bibitem{Kundu:2023hbk}
A.~Kundu, V.~Malvimat and R.~Sinha,
JHEP \textbf{09} (2023), 011
doi:10.1007/JHEP09(2023)011
[arXiv:2303.03426 [hep-th]].



\bibitem{He:2024hkw}
P.~Z.~He and H.~Q.~Zhang,
JHEP \textbf{03} (2025), 142
doi:10.1007/JHEP03(2025)142
[arXiv:2411.16302 [hep-th]].





\bibitem{He:2025guu}
P.~Z.~He, L.~H.~Liu, H.~Q.~Zhang and Q.~Q.~Jiang,
[arXiv:2509.14742 [hep-th]].

\bibitem{He:2024xjp}
P.~Z.~He and H.~Q.~Zhang,
JHEP \textbf{11} (2024), 014
doi:10.1007/JHEP11(2024)014
[arXiv:2407.02756 [hep-th]].


\bibitem{Caputa:2022eye}
P.~Caputa and S.~Liu,
Phys. Rev. B \textbf{106} (2022) no.19, 195125
doi:10.1103/PhysRevB.106.195125
[arXiv:2205.05688 [hep-th]].

\bibitem{Camargo:2022rnt}
H.~A.~Camargo, V.~Jahnke, K.~Y.~Kim and M.~Nishida,
JHEP \textbf{05} (2023), 226
doi:10.1007/JHEP05(2023)226
[arXiv:2212.14702 [hep-th]].





\bibitem{Bhattacharyya:2025lsc}
A.~Bhattacharyya, S.~Gool and S.~S.~Haque,
[arXiv:2509.14810 [hep-th]].



\bibitem{Zhai:2024tkz}
K.~H.~Zhai, L.~H.~Liu and H.~Q.~Zhang,
[arXiv:2412.08925 [hep-th]].


\bibitem{Carolan:2024wov}
E.~Carolan, A.~Kiely, S.~Campbell and S.~Deffner,
EPL \textbf{147} (2024) no.3, 38002
doi:10.1209/0295-5075/ad5b17
[arXiv:2404.03529 [quant-ph]].

\bibitem{Bhattacharya:2023zqt}
A.~Bhattacharya, P.~Nandy, P.~P.~Nath and H.~Sahu,
JHEP \textbf{12} (2023), 066
doi:10.1007/JHEP12(2023)066
[arXiv:2303.04175 [quant-ph]].


\bibitem{Liu:2022god}
C.~Liu, H.~Tang and H.~Zhai,
Phys. Rev. Res. \textbf{5} (2023) no.3, 033085
doi:10.1103/PhysRevResearch.5.033085
[arXiv:2207.13603 [cond-mat.str-el]].


\bibitem{Bhattacharjee:2022lzy}
B.~Bhattacharjee, X.~Cao, P.~Nandy and T.~Pathak,
JHEP \textbf{03} (2023), 054
doi:10.1007/JHEP03(2023)054
[arXiv:2212.06180 [quant-ph]].




\bibitem{Li:2025fqz}
Z.~Li and J.~Tian,
[arXiv:2506.13481 [hep-th]].

\bibitem{Aguilar-Gutierrez:2025kmw}
S.~E.~Aguilar-Gutierrez, H.~A.~Camargo, V.~Jahnke, K.~Y.~Kim and M.~Nishida,
[arXiv:2506.03273 [hep-th]].


\bibitem{Heller:2024ldz}
M.~P.~Heller, J.~Papalini and T.~Schuhmann,
[arXiv:2412.17785 [hep-th]].


\bibitem{Das:2024tnw}
R.~N.~Das, S.~Demulder, J.~Erdmenger and C.~Northe,
JHEP \textbf{06} (2025), 166
doi:10.1007/JHEP06(2025)166
[arXiv:2412.09673 [hep-th]].


\bibitem{Rabinovici:2023yex}
E.~Rabinovici, A.~S{\'a}nchez-Garrido, R.~Shir and J.~Sonner,
JHEP \textbf{08} (2023), 213
doi:10.1007/JHEP08(2023)213
[arXiv:2305.04355 [hep-th]].



\bibitem{Bhattacharyya:2025gvd}
A.~Bhattacharyya, S.~Ghosh, S.~Pal and A.~Vinod,
[arXiv:2502.13208 [hep-th]].





\bibitem{Ambrosini:2024sre}
M.~Ambrosini, E.~Rabinovici, A.~S{\'a}nchez-Garrido, R.~Shir and J.~Sonner,
JHEP \textbf{08} (2025), 059
doi:10.1007/JHEP08(2025)059
[arXiv:2412.15318 [hep-th]].



\bibitem{Chattopadhyay:2023fob}
A.~Chattopadhyay, A.~Mitra and H.~J.~R.~van Zyl,
Phys. Rev. D \textbf{108} (2023) no.2, 025013
doi:10.1103/PhysRevD.108.025013
[arXiv:2302.10489 [hep-th]].


\bibitem{Rabinovici:2025otw}
E.~Rabinovici, A.~S{\'a}nchez-Garrido, R.~Shir and J.~Sonner,
[arXiv:2507.06286 [hep-th]].


\bibitem{Nandy:2024evd}
P.~Nandy, A.~S.~Matsoukas-Roubeas, P.~Mart{\'\i}nez-Azcona, A.~Dymarsky and A.~del Campo,
Phys. Rept. \textbf{1125-1128} (2025), 1-82
doi:10.1016/j.physrep.2025.05.001
[arXiv:2405.09628 [quant-ph]].



\bibitem{Wu:2022xwy}
S.~M.~Wu and H.~S.~Zeng,
Eur. Phys. J. C \textbf{82} (2022) no.1, 4
doi:10.1140/epjc/s10052-021-09954-4
[arXiv:2201.02333 [quant-ph]].


\bibitem{Wu:2023spa}
S.~M.~Wu, X.~W.~Teng, J.~X.~Li, S.~H.~Li, T.~H.~Liu and J.~C.~Wang,
Phys. Lett. B \textbf{848} (2024), 138334
doi:10.1016/j.physletb.2023.138334
[arXiv:2311.12362 [gr-qc]].

\bibitem{Wu:2023sye}
S.~M.~Wu, X.~W.~Fan, R.~D.~Wang, H.~Y.~Wu, X.~L.~Huang and H.~S.~Zeng,
JHEP \textbf{11} (2023), 232
doi:10.1007/JHEP11(2023)232
[arXiv:2304.00984 [gr-qc]].


\bibitem{Li:2025bzd}
S.~H.~Li, S.~H.~Shang and S.~M.~Wu,
JHEP \textbf{05} (2025), 214
doi:10.1007/JHEP05(2025)214
[arXiv:2502.05881 [gr-qc]].

\bibitem{Li:2025jlu}
W.~M.~Li, J.~Lu and S.~M.~Wu,
[arXiv:2505.07476 [gr-qc]].

\bibitem{Liu:2024wpa}
W.~Liu, C.~Wen and J.~Wang,
JHEP \textbf{01} (2025), 184
doi:10.1007/JHEP01(2025)184
[arXiv:2410.21681 [gr-qc]].

\bibitem{Liu:2025hcx}
X.~Liu, W.~Liu and S.~M.~Wu,
[arXiv:2511.12245 [gr-qc]].



\bibitem{Choudhury:2020hil}
S.~Choudhury, S.~Chowdhury, N.~Gupta, A.~Mishara, S.~P.~Selvam, S.~Panda, G.~D.~Pasquino, C.~Singha and A.~Swain,
Symmetry \textbf{13} (2021) no.7, 1301
doi:10.3390/sym13071301
[arXiv:2012.10234 [hep-th]].


\bibitem{Bhargava:2020fhl}
P.~Bhargava, S.~Choudhury, S.~Chowdhury, A.~Mishara, S.~P.~Selvam, S.~Panda and G.~D.~Pasquino,
SciPost Phys. Core \textbf{4} (2021), 026
doi:10.21468/SciPostPhysCore.4.4.026
[arXiv:2009.03893 [hep-th]].


\bibitem{Lehners:2020pem}
J.~L.~Lehners and J.~Quintin,
Phys. Rev. D \textbf{103} (2021) no.6, 063527
doi:10.1103/PhysRevD.103.063527
[arXiv:2012.04911 [hep-th]].


\bibitem{Bhattacharyya:2020rpy}
A.~Bhattacharyya, S.~Das, S.~Shajidul Haque and B.~Underwood,
Phys. Rev. D \textbf{101} (2020) no.10, 106020
doi:10.1103/PhysRevD.101.106020
[arXiv:2001.08664 [hep-th]].


\bibitem{Adhikari:2021ked}
K.~Adhikari, S.~Choudhury, H.~N.~Pandya and R.~Srivastava,
Symmetry \textbf{15} (2023) no.3, 664
doi:10.3390/sym15030664
[arXiv:2108.10334 [gr-qc]].









\bibitem{Adhikari:2022oxr}
K.~Adhikari and S.~Choudhury,
Fortsch. Phys. \textbf{70} (2022) no.12, 2200126
doi:10.1002/prop.202200126
[arXiv:2203.14330 [hep-th]].




\bibitem{Li:2024kfm}
T.~Li and L.~H.~Liu,
JHEP \textbf{04} (2024), 123
doi:10.1007/JHEP04(2024)123
[arXiv:2401.09307 [hep-th]].



\bibitem{Li:2024ljz}
T.~Li and L.~H.~Liu,
[arXiv:2408.03293 [hep-th]].


\bibitem{Li:2024iji}
T.~Li and L.~H.~Liu,
[arXiv:2405.01433 [hep-th]].


\bibitem{Zhai:2024odw}
K.~H.~Zhai and L.~H.~Liu,
[arXiv:2411.18405 [hep-th]].

\bibitem{Baumann:2009ds}
D.~Baumann,
doi:10.1142/9789814327183{\_}0010
[arXiv:0907.5424 [hep-th]].


\bibitem{Armendariz-Picon:1999hyi}
C.~Armendariz-Picon, T.~Damour and V.~F.~Mukhanov,
Phys. Lett. B \textbf{458} (1999), 209-218
doi:10.1016/S0370-2693(99)00603-6
[arXiv:hep-th/9904075 [hep-th]].


\bibitem{Garriga:1999vw}
J.~Garriga and V.~F.~Mukhanov,
Phys. Lett. B \textbf{458} (1999), 219-225
doi:10.1016/S0370-2693(99)00602-4
[arXiv:hep-th/9904176 [hep-th]].

\bibitem{Alishahiha:2004eh}
M.~Alishahiha, E.~Silverstein and D.~Tong,
Phys. Rev. D \textbf{70} (2004), 123505
doi:10.1103/PhysRevD.70.123505
[arXiv:hep-th/0404084 [hep-th]].


\bibitem{Silverstein:2003hf}
E.~Silverstein and D.~Tong,
Phys. Rev. D \textbf{70} (2004), 103505
doi:10.1103/PhysRevD.70.103505
[arXiv:hep-th/0310221 [hep-th]].


\bibitem{Chen:2020uhe}
C.~Chen, X.~H.~Ma and Y.~F.~Cai,
Phys. Rev. D \textbf{102} (2020) no.6, 063526
doi:10.1103/PhysRevD.102.063526
[arXiv:2003.03821 [astro-ph.CO]].

\bibitem{Achucarro:2012sm}
A.~Achucarro, J.~O.~Gong, S.~Hardeman, G.~A.~Palma and S.~P.~Patil,
JHEP \textbf{05} (2012), 066
doi:10.1007/JHEP05(2012)066
[arXiv:1201.6342 [hep-th]].


\bibitem{Cai:2018tuh}
Y.~F.~Cai, X.~Tong, D.~G.~Wang and S.~F.~Yan,
Phys. Rev. Lett. \textbf{121} (2018) no.8, 081306
doi:10.1103/PhysRevLett.121.081306
[arXiv:1805.03639 [astro-ph.CO]].


\bibitem{Chen:2020uhe}
C.~Chen, X.~H.~Ma and Y.~F.~Cai,
Phys. Rev. D \textbf{102} (2020) no.6, 063526
doi:10.1103/PhysRevD.102.063526
[arXiv:2003.03821 [astro-ph.CO]].


\bibitem{Chen:2019zza}
C.~Chen and Y.~F.~Cai,
JCAP \textbf{10} (2019), 068
doi:10.1088/1475-7516/2019/10/068
[arXiv:1908.03942 [astro-ph.CO]].


\bibitem{Bhattacharya:2022gbz}
A.~Bhattacharya, P.~Nandy, P.~P.~Nath and H.~Sahu,
JHEP \textbf{12} (2022), 081
doi:10.1007/JHEP12(2022)081
[arXiv:2207.05347 [quant-ph]].









\bibitem{Hetyei_2009}
H. G{\'a}bor, 
Proceedings of the Royal Society A: Mathematical, Physical and Engineering Sciences, \textbf{10} (2009), no.2117, 1471-2946,
doi:10.1098/rspa.2009.0497
[arXiv:0909.4352[Mathematics-Quantum Algebra]].




\bibitem{Nizami:2023dkf}
A.~A.~Nizami and A.~W.~Shrestha,
Phys. Rev. E \textbf{108} (2023) no.5, 5
doi:10.1103/PhysRevE.108.054222
[arXiv:2305.00256 [quant-ph]].














\bibitem{Zhai:2025abc}
K.~H.~Zhai, L.~H.~Liu and H.~Q.~Zhang,
[arXiv:2505.20595 [quant-ph]].


\bibitem{Liu:2019xhn}
L.~H.~Liu and W.~L.~Xu,
Chin. Phys. C \textbf{44} (2020) no.8, 085103
doi:10.1088/1674-1137/44/8/085103
[arXiv:1911.10542 [astro-ph.CO]].


\bibitem{Liu:2020zzv}
L.~H.~Liu and T.~Prokopec,
JCAP \textbf{06} (2021), 033
doi:10.1088/1475-7516/2021/06/033
[arXiv:2005.11069 [astro-ph.CO]].


\bibitem{Zhang:2022bde}
X.~z.~Zhang, L.~h.~Liu and T.~Qiu,
Phys. Rev. D \textbf{107} (2023) no.4, 043510
doi:10.1103/PhysRevD.107.043510
[arXiv:2207.07873 [hep-th]].


\bibitem{Liu:2021rgq}
L.~H.~Liu,
Chin. Phys. C \textbf{47} (2023) no.1, 015105
doi:10.1088/1674-1137/ac9d28
[arXiv:2107.07310 [astro-ph.CO]].


\bibitem{Liu:2020zlr}
L.~H.~Liu, B.~Liang, Y.~C.~Zhou, X.~D.~Liu, W.~L.~Xu and A.~C.~Li,
Phys. Rev. D \textbf{103} (2021) no.6, 063515
doi:10.1103/PhysRevD.103.063515
[arXiv:2007.08278 [astro-ph.CO]].


\bibitem{Liu:2018hno}
L.~H.~Liu, T.~Prokopec and A.~A.~Starobinsky,
Phys. Rev. D \textbf{98} (2018) no.4, 043505
doi:10.1103/PhysRevD.98.043505
[arXiv:1806.05407 [gr-qc]].



\bibitem{Liu:2018htf}
L.~H.~Liu,
doi:10.1007/s10773-018-3809-0
[arXiv:1807.00666 [gr-qc]].



















	
	




\end{thebibliography}
\end{document}